\documentclass[12pt]{article}
\usepackage{amsfonts}
\usepackage{amsmath}
\usepackage{amssymb}
\usepackage{array}
\usepackage{bigints}
\usepackage{booktabs}
\usepackage[nosort]{cite}
\usepackage{color}
\usepackage{dsfont}
\usepackage{float}
\usepackage{framed}
\usepackage{graphicx}
\usepackage{indentfirst}
\usepackage{mathrsfs}
\usepackage{multirow}
\usepackage{pdflscape}
\usepackage{setspace}
\usepackage{subdepth}
\usepackage{subfig}
\usepackage{titlesec}
\usepackage[dotinlabels]{titletoc}
\usepackage{wrapfig}
\usepackage[all]{xy}
\usepackage{young}
\usepackage[vcentermath]{youngtab}
\usepackage{relsize}
\usepackage{stackengine}
\usepackage{datetime}

\usepackage{hyperref}

\numberwithin{equation}{section}

\usepackage{verbatim}

\newcommand{\be}{\begin{equation}}
\newcommand{\ee}{\end{equation}}

\def\({\left(} \def\){\right)}
\def\[{\left[} \def\]{\right]}

\def\sgn{\text{sgn}}

\def\mF{\mathcal{F}}
\def\mS{\mathcal{S}}
\def\mFa{\mathcal{F}_{\text{amp}}}

\def\mO{\mathcal{O}}
\def\mC{\mathcal{C}}

\def\mI{\mathcal{I}}

\def\eps{\epsilon}

\newcommand{\bea}{\begin{eqnarray}}
\newcommand{\eea}{\end{eqnarray}}

\newcommand{\bml}{\begin{multline}}
\newcommand{\emll}{\end{multline}}

\usepackage[left=2.5cm,right=2.5cm,top=2.5cm,bottom=2.5cm]{geometry}
\titleformat{\section}{\normalfont\bfseries}{\thesection.}{4pt}{}
\titlespacing{\section}{0pt}{22pt}{6pt}

\titleformat{\subsection}{\normalfont\itshape}{\thesubsection.}{4pt}{}
\titlespacing{\subsection}{0pt}{18pt}{6pt}

\titleformat{\subsubsection}{\normalfont\itshape}{\thesubsubsection.}{4pt}{}
\titlespacing{\subsubsection}{0pt}{16pt}{6pt}

\def\ie{\begin{equation}\begin{aligned}}
\def\fe{\end{aligned}\end{equation}}

\def\ep{\varepsilon}
\def\1{{\mathds 1}}

\DeclareFontShape{OT1}{cmr}{mx}{n}%
    {<->cmr10}{}
\newcommand{\mytitlefont}{\fontseries{mx}\selectfont}
\DeclareMathAlphabet{\titlemath}{OT1}{cmr}{mx}{n}

\begin{document}

% TITLEPAGE

\begin{titlepage}

\begin{center}

~\\[2cm]

{\fontsize{20pt}{0pt} \mytitlefont The Bulk Dual of SYK: Cubic Couplings}

~\\[0.5cm]

{\fontsize{14pt}{0pt} David J.~Gross and Vladimir Rosenhaus}

~\\[0.1cm]

\it{Kavli Institute for Theoretical Physics}\\ \it{University of California, Santa Barbara, CA 93106}

~\\[0.8cm]

\end{center}

\noindent The SYK model, a  quantum mechanical model of $N \gg 1$ Majorana fermions $\chi_i$, with a $q$-body, random interaction, is a novel realization of holography. It is known that the AdS$_2$ dual contains a tower of massive particles, yet there is at present no proposal for the bulk theory. As SYK is solvable in the $1/N$ expansion, one can systematically derive the bulk. We initiate such a  program, by  analyzing  the fermion two, four and six-point functions, from which we extract  the tower of singlet, large $N$ dominant, operators, their dimensions,  and their three-point correlation functions. These determine the masses of the bulk fields and their cubic couplings. We present these couplings, analyze their structure and discuss the simplifications that arise for large $q$.

\vfill 

%\begin{flushleft}
%September 2016
%\end{flushleft}

\end{titlepage}

% TABLE OF CONTENTS

\tableofcontents

~\\[.1cm]
\newpage

\section{Introduction}

The SYK model \cite{SY, Kitaev} is a $0+1$ dimensional theory of $N \gg 1$ Majorana fermions $\chi_i$, with a $q$-body, random interaction. This model is of great interest for the following reasons. First, in the infrared, the model is  ``nearly'' conformally invariant \cite{SY}, even more it is  ``nearly'' diffeomorphism  invariant \cite{Kitaev}; second, the model has the same Lyapunov exponent as a black hole\cite{Kitaev}; both features suggesting the existence of a gravitational dual in AdS$_2$. The nature of the bulk dual is largely unknown, except that it likely contains AdS$_2$ dilaton gravity \cite{AP}. But, in addition to the gravitational degrees of freedom, the bulk model must contain a tower of massive particles dual to the tower of  bilinear, primary, $O(N)$ singlet operators in SYK,  schematically of the form, $\mO_n = \sum_{i=1}^N \chi_i \partial_{\tau}^{1+2n} \chi_i$ \cite{MS, Kitaev, PR}. The standard AdS/CFT mapping dictates that each of these $\mO_n$ is dual to a bulk scalar field $\phi_n$. The  mass $m_n$  of each $\phi_n$ is related to the infrared dimension $h_n$ of $\mO_n$ by $m_n^2 = h_n(h_n-1)$ (in units where the AdS radius is one). Our goal is to construct this dual theory.

SYK   can be regarded as being  somewhere in between a vector model and a matrix model. Like a vector model, SYK has an $O(N)$ symmetry, with fields in the fundamental representation. Yet, the physical properties of SYK are closer to, for instance, strongly coupled $\mathcal{N}=4$ Yang-Mills: SYK is  maximally chaotic \cite{Kitaev} and appears to display features of random matrix theories  \cite{Cotler:2016fpe}.
The SYK model is currently in a different state from both the  gauge theory and the $O(N)$ theory examples of AdS/CFT. There is no known brane construction of SYK, nor is there even a candidate for what the bulk theory could be. However, the solvability of SYK at large $N$ gives us, in principle,  the ability to directly derive the bulk.

The canonical example of AdS/CFT is the duality between four-dimensional supersymmetric $\mathcal{N} =4$ Yang-Mills and Type IIB string theory in AdS$_5 \times S^5$. This duality was conjectured  on the basis of consideration of the near horizon limit of $D3$ branes, with closed string theory living in the bulk, and Yang-Mills living on the branes \cite{Maldacena:1997re}. There have been numerous tests of the duality. One of the earliest  was the matching of the three-point function of chiral primary operators in $\mathcal{N}=4$ with the corresponding bulk calculation, arising from the cubic couplings of the supergravity fields compactified on $S^5$ \cite{ Lee:1998bxa, Freedman:1998tz}. Computations in strongly coupled $\mathcal{N}=4$ are generally intractable; but in this particular case, supersymmetric nonrenormalization theorems made it possible to perform the computation in the free field limit. 

The duality between $\mathcal{N}=4$ Yang-Mills and string theory takes advantage of the fact that the coupling of  $\mathcal{N} =4$ is marginal and in the strong coupling limit only a few low dimension operators survive, which are dual to the few local supergravity fields that survive in the low energy limit of string theory. In principle one could imagine using the duality to derive the full content of supergravity in the bulk from the strongly coupled $\mathcal{N} =4$ Yang-Mills theory. Even more, one could in principle use the duality to construct the full content of classical string theory in AdS$_5 \times S^5$ from the large $N$ super Yang-Mills theory. 
SYK, unlike $\mathcal{N}=4$, lacks a marginal coupling and there is no  limit in  which only a finite number of operators of finite dimensions survive. Neither adding flavor to SYK \cite{GR} nor supersymmetry \cite{Fu:2016vas} alters this conclusion. Thus, at the very least, the  dual theory to SYK must involve an infinite number of local fields, and could very well be described by string theory or some other theory of extended objects.

Another canonical class of AdS/CFT dualities is that between the  $O(N)$ vector model and higher spin theory \cite{Klebanov:2002ja, Vasiliev:1999ba}. The bulk/boundary matching of the three-point functions of the $O(N)$ singlets was performed in \cite{Giombi:2009wh}. This example is perhaps more analogous to SYK, as the bulk theory contains an infinite number of fields. SYK may be simpler however, as the bulk fields are massive.

\begin{figure}[t]
\centering
\subfloat[]{
\includegraphics[width=1.9in]{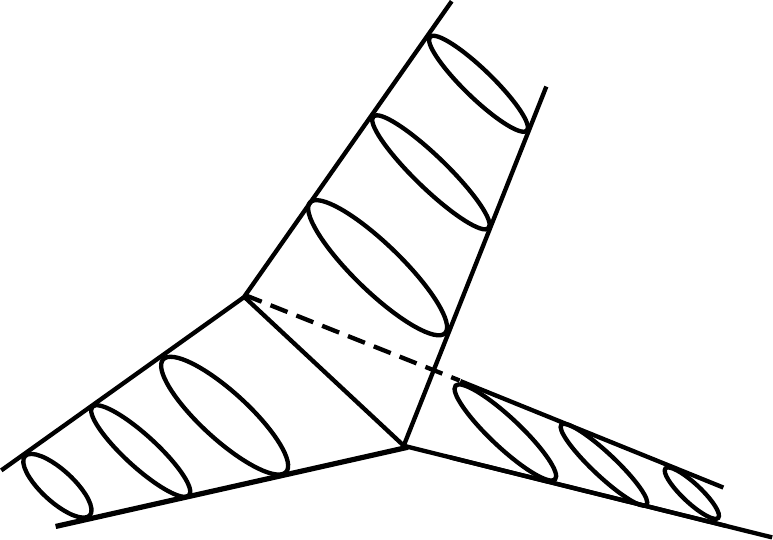} 
} \ \ \ \ \ \ \ \ \ \ \ \ \ \ \ \ \
\subfloat[]{
\includegraphics[width=2in]{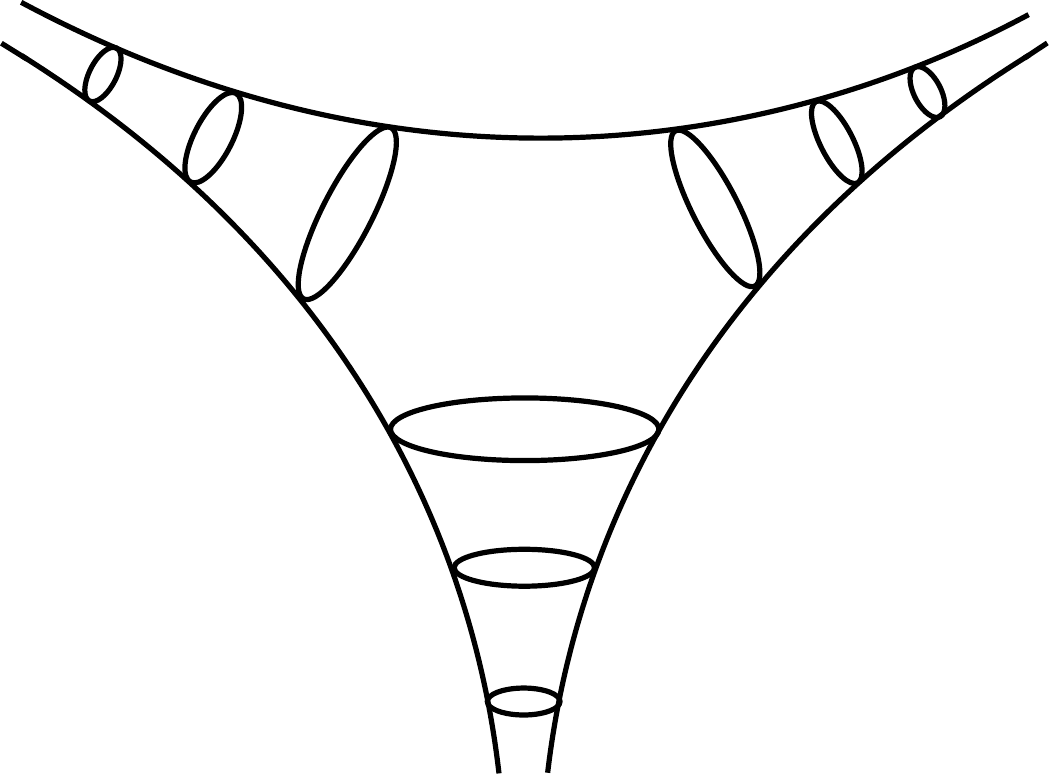}
}
\caption{The two classes of Feynman diagrams relevant for the fermion six-point function.} \label{FigIntro3}
\end{figure}

To leading order in $1/N$  
the bulk fields, $\phi_n$,  are free,   with higher-point couplings suppressed by powers of $1/\sqrt{N}$. 
The lowest order interaction term is  given by a cubic coupling, $\frac{1}{\sqrt{N}}\lambda_{n m k } \phi_n \phi_m \phi_k$. In this paper we compute $\lambda_{n m k }$ by studying  the fermion six-point function. 

In Sec.~\ref{sec:4pt} we review the fermion four-point function, $N^{-2} \sum_{i, j} \langle\chi_i (\tau_1) \chi_i(\tau_2) \chi_j(\tau_3) \chi_j(\tau_4)\rangle$. In the short time limit, $\tau_1 \rightarrow \tau_2,\,\, \,  \tau_3\rightarrow \tau_4$, and in the infrared, $|\tau_2 - \tau_4| \rightarrow \infty$, this reduces to a sum over the conformal two-point functions of the bilinear operators,  $\langle \mO_n(\tau_2) \mO_n(\tau_4) \rangle$, from which we extract the masses of the bulk fields.  In Sec.~\ref{sec:6pt} we study the fermion six-point function $N^{-3} \sum_{i, j, l} \langle\chi_i (\tau_1) \chi_i(\tau_2) \chi_j(\tau_3) \chi_j(\tau_4) \chi_l (\tau_5) \chi_l(\tau_6)\rangle$. In the short time limit, $\tau_1 \rightarrow \tau_2,\,\,\,  \tau_3\rightarrow \tau_4,\,\,\, \tau_5\rightarrow \tau_6$, and in the infrared, this reduces to a sum over the conformal three-point functions, $\langle \mO_n(\tau_2) \mO_m(\tau_4) \mO_k(\tau_6)\rangle$. There are two classes of Feynman diagrams that contribute to the six-point function at order $1/N^2$, as shown in Fig.~\ref{FigIntro3}. We will refer to these as the ``contact'' and planar diagrams, respectively. These  diagrams can be summed as  the six-point function can be compactly written as three four-point functions glued together. The three-point functions $\langle \mO_n(\tau_2) \mO_m(\tau_4) \mO_k(\tau_6)\rangle$ are determined by conformal invariance up to a constant,  $c_{ n m k}$, which we extract from the  calculation above.

\begin{figure}[t]
\centering
\includegraphics[width=1.3in]{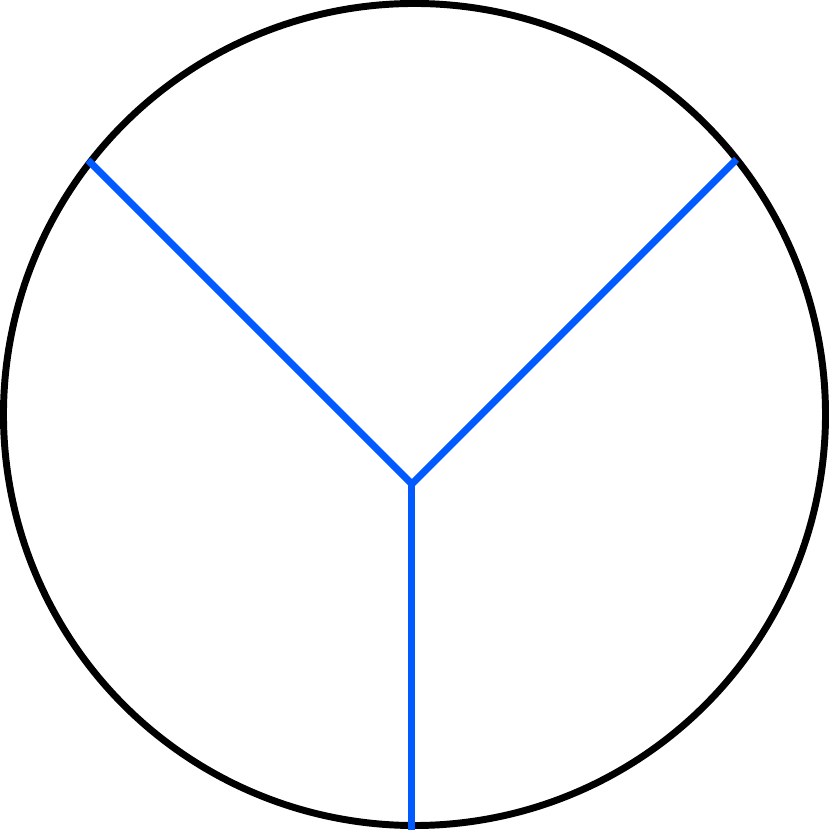} 
\caption{A cubic coupling of the bulk fields $\phi_n$ gives rise to the tree level interaction of the bulk fields shown in this Feynman-Witten diagram. Extrapolating the bulk points to the boundary gives, via the AdS/CFT dictionary, the CFT  three-point function of the composites $\mO_n$. } \label{FigIntroW}
\end{figure}

A cubic interaction in the bulk gives rise to the tree-level Witten diagram shown in Fig.~\ref{FigIntroW}.
 Relating the couplings  $\lambda_{n m k }$ of this cubic interaction to  $c_{ n m k}$ is straightforward, and is discussed in Sec.~\ref{sec:bulk}. The expressions for $\lambda_{n m k }$  are quite complicated, but simplify dramatically when $q \to \infty$. In this  limit we can derive  an explicit analytic expression for  $\lambda_{n m k }$ in the large $q$ limit.

 There are two contributions to $\lambda_{n m k}$. The first, $\lambda_{n m k}^{(1)}$,  due to the ``contact'' diagrams,  can be written explicitly for any $q$, takes a simple form when $q \to \infty$ and decays with large $n, m, k$. The second, $\lambda_{n m k}^{(2)}$, arising from the planar diagrams, is dominant at large $n, m, k$ and is related in a simple way to the bulk couplings one would have for the bulk dual of a generalized free field theory, which is studied in Appendix~\ref{ap:free}.
 
 Finally, in Sec.~\ref{sec:dis} we discuss the final results and speculate on their implications for the complete dual theory.

\section{A Tower of Particles} \label{sec:4pt}
This section largely reviews aspects of SYK that we will need later. In Sec.~\ref{sec:two} we recall the construction of the (large $N$) two-point function of the fermions  in the infrared or strong coupling limit. In this limit SYK is almost conformally invariant, with the fermions having dimension $\Delta = 1/q$ for the model with a $q$-body interaction. In Sec.~\ref{sec:four} we study the fermion four-point function, which is essentially given by a sum of conformal blocks of the bilinears of fermions, $\mO_n$, having dimension $h_n$. In the large $q$ limit these dimensions take the simple form, $h_n = 2n + 1+ 2\eps_n$, where $\eps_n$ scales as $1/q$. In Sec.~\ref{sec:OPE}  we take the double short time limit of the fermion four-point function, turning it into a sum of two-point functions of the bilinears $\mO_n$, and in  Sec.~\ref{sec:bulkpre} we use these results to determine the quadratic part of the massive scalar bulk Lagrangian.

\subsection{Fermion two-point function} \label{sec:two}
\begin{figure}[t]
\centering
\includegraphics[width=4.5in]{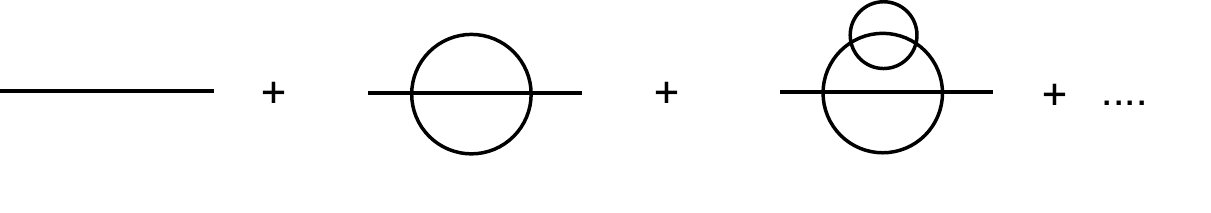}
\caption{The fermion two-point function, to leading order in $1/N$, is a sum of melon diagrams. This figure, as well as all figures, are for $q=4$ SYK.  In addition, we have suppressed the dashed lines commonly drawn to indicate   $\langle J_{i_1\ldots i_q} J_{i_1 \ldots i_q}\rangle$ contractions. } \label{Figmelon}
\end{figure}

The SYK model \cite{Kitaev}, a cousin of the SY model \cite{SY, PGKS}, describes $N$ Majorana fermions satisfying $\{\chi_i, \chi_j\} = \delta_{i j}$, with action:
\be \label{SYK}
S= \int d \tau\,  \( \frac{1}{2} \sum_{i=1}^N\chi_i \frac{d }{d \tau} \chi_i\, +\, \frac{(i)^{\frac{q}{2}}}{q!} \sum_{i_1, \ldots, i_q=1}^N J_{ i_1 i_2 \ldots i_q} \chi_{i_1} \chi_{i_2}\cdots \chi_{i_q} \)~,
\ee
where the coupling $J_{i_1, \ldots, i_q}$ is totally antisymmetric and, for each $i_1, \ldots, i_q$, is chosen from a Gaussian ensemble, with variance,
\be \label{disA}
\frac{1}{(q-1)!} \sum_{i_2, \ldots, i_q=1}^{N}\langle J_{i_1 i_2 \ldots i_q}  J_{i_1 i_2 \ldots i_q}\rangle= J^2~.
\ee
One can consider SYK for any even $q\geq 2$, with  $q=4$ being the prototypical case.
Major simplification occurs as $q \to \infty $, a limit which we will exploit in the following.

At zero coupling, the time-ordered Euclidean two-point function,  $\langle \chi_i(\tau) \chi_j(0)\rangle \equiv G(\tau) \delta_{i j}$, is given by,
\be \label{Gweak}
G_0(\tau) = \frac{1}{2} \sgn(\tau)~,
\ee
where the factor $\sgn(\tau)$ ($\sgn(\tau) = 1$ for $\tau>0$ and $\sgn(\tau) = -1$ for $\tau<0$) accounts for the fermion anticommutation. The two-point function at strong coupling, to leading order in $1/N$, is found by summing all melon diagrams, such as the ones shown in Fig.~\ref{Figmelon}. The result, for $|J \tau| \gg 1$, is \cite{SY, Kitaev},
\be \label{Gstrong}
G(\tau) = b \frac{\sgn(\tau)}{|J \tau|^{2\Delta}}~,
\ee
where 
\be \label{DeltaSYK}
\Delta= \frac{1}{q}~, \ \ \ \ \ \ \ \ b^{q} =\frac{1}{2\pi}\(1 - 2 \Delta\) \tan \pi \Delta~.
\ee

The two-point function, as well as all other correlation functions we compute, are averaged over the disorder, which restores $O(N)$ invariance.~\footnote{One can consider variants of SYK that do not have disorder. The simplest is to make the couplings $J_{i_1 \ldots i_q}$ be nearly static quantum variables \cite{MPRS}. A better way of eliminating disorder is to turn SYK into a tensor model \cite{Witten:2016iux} (see also, \cite{Klebanov:2016xxf, Peng:2016mxj}), analogous to the ones previously studied \cite{Sasakura:1990fs}. To leading order in $1/N$ all these approaches agree.} In the limit of large $q$, the IR fermion dimension ($\Delta$) approaches the vanishing UV dimension,  allowing one to find a simple analytic expression for the two-point function at all energies \cite{MS}.  One way to understand why this is possible is because at large $q$ one only needs to sum a particular subset of melon diagrams, due to their  combinatorial dominance \cite{GR}.

\subsection{Fermion four-point function } \label{sec:four}
\begin{figure}[t]
\centering
\subfloat[]{
\includegraphics[width=5in]{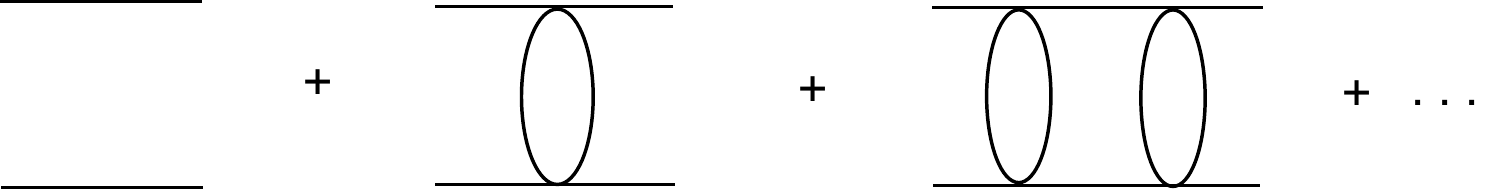}
}\\
\subfloat[]{\!\!\!\!\!\!\!\!\!\!\!\!\!\!\!\!\!\!\!\!\!\!\!\!
\includegraphics[width=2.5in]{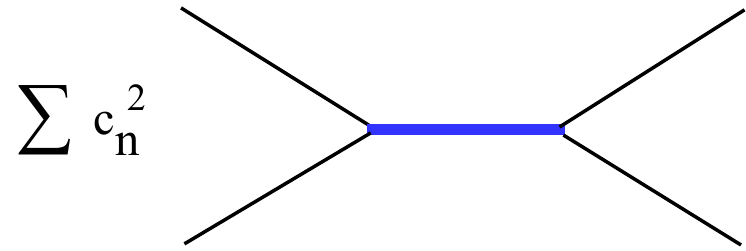}
}
\caption{(a) The fermion four-point function, at order $1/N$, is a sum of ladder diagrams. (b) It can be represented as a sum of conformal blocks, involving the $\mO_n$.} \label{Figladder}
\end{figure} 

In large $N$ vector models, the leading connected part of a $2k$-point correlation function scales as $1/N^{k-1}$ and is completely determined by linear integral equations whose kernel is expressed in terms of the two-point function. The SYK four-point function to order $1/N$, studied in \cite{Kitaev, PR, MS}, is given by,
\be \label{4ptSum}
\frac{1}{N^2}\sum_{i, j=1}^N\langle \chi_i(\tau_1) \chi_i(\tau_2) \chi_j(\tau_3) \chi_j(\tau_4)\rangle = G(\tau_{12}) G(\tau_{34}) + \frac{1}{N}\mF(\tau_1, \tau_2, \tau_3, \tau_4)~,
\ee
where $\tau_{12} \equiv \tau_1 - \tau_2$ and $\mF$ is given by the sum of ladder diagrams, as shown in  Fig.~\ref{Figladder}. Due to the restored $O(N)$ invariance the leading behavior in $1/N$ is completely captured by $\mF$.

Let us briefly recall how $\mF$ is computed. 
 The first diagram in Fig.~\ref{Figladder}, although disconnected, is suppressed by $1/N$ as it requires setting the indices to be equal, $i=j$. We denote this diagram by $\mF_0$,~\footnote{In Fig~\ref{Figladder}, as well as elsewhere, we have not drawn the crossed diagrams, i.e., the first diagram in Fig.~\ref{Figladder} only corresponds to the first term in $\mF_0$.}
\be
\mF_0 = - G(\tau_{13}) G(\tau_{24})+ G(\tau_{14}) G(\tau_{23})~.
\ee
Now we must sum all the other diagrams. Letting $K$ denote the kernel that adds a rung to the ladder, then summing the ladders yields, $\mF =(1 + K + K^2 + \ldots) \mF_0 = (1- K)^{-1} \mF_0$.  The difficult step  is inverting $1-K$.

The result of this computation  is that, in the infrared, the four-point function can (almost) be written as a sum of conformal blocks of operators of dimensions $h_n$. The reason we say almost is because there is one block, arising from an operator of dimension $h=2$, that breaks conformal invariance. We will discuss it shortly; for now we separate the four-point function into the conformally invariant part $\mF_c$ and the $h=2$ part $\mF_{h=2}$, 
\be \label{Fc2}
\mF = \mF_c + \mF_{h=2}~,
\ee 
where a convenient form for $\mF_c$ is \cite{MS}, 
\be \label{4pt}
\mF_c(\tau_1, \tau_2, \tau_3, \tau_4)  =  G(\tau_{12}) G(\tau_{34})\sum_{n=1}^{\infty} c_n^2\, x^{h_n}\, {_2} F_1(h_n,h_n, 2 h_n, x)~,
\ee
where $x$ is the conformal cross-ratio of the four times $\tau_i$, 
\be
x = \frac{\tau_{12} \tau_{34}}{\tau_{13} \tau_{24}}~,
\ee
and $h_n$ are the values of $h$ for which the eigenvalue of the kernel operator  is equal to one, $k_c(h_n) = 1$, where,
\be \label{kch}
k_c(h) = -(q-1) \frac{\Gamma\(\frac{3}{2} - \frac{1}{q}\) \Gamma\(1 - \frac{1}{q}\) }{\Gamma\(\frac{1}{2} + \frac{1}{q}\) \Gamma\( \frac{1}{q}\)} \frac{ \Gamma\(\frac{1}{q} + \frac{h}{2}\) \Gamma\(\frac{1}{2} + \frac{1}{q} - \frac{h}{2}\) }{   \Gamma\(\frac{3}{2} - \frac{1}{q} - \frac{h}{2}\) \Gamma\(1- \frac{1}{q} + \frac{h}{2}\) }
~.
\ee
The kernel always has an $h=2$ eigenvalue, as $k_c(h=2) = 1$ for all values of $q$. The sum in (\ref{4pt}) contains the rest of the terms  with  $h_n>2$, as the $h=2$ contribution is accounted for by the nonconformal piece $\mF_{h=2}$ in (\ref{Fc2}). The constants $c_n$ in (\ref{4pt}) are the OPE coefficients, 
\be \label{opexi}
\frac{1}{N}\sum_{i=1}^N\chi_i(\tau_1) \chi_i(\tau_2) = \frac{1}{\sqrt{N}}\sum_n c_n  {\sgn(\tau_{12})\over |\tau_{12}|^{2\Delta -h_n} }\mO_n\({\tau_1+\tau_2 \over 2}\)\,,
\ee
\be \label{cm}
c_n^2 = \frac{2 q}{(q-1) (q-2) \tan \frac{\pi}{q}} \frac{(h_n - 1/2)}{ \tan( \pi h_n/2)} \frac{\Gamma(h_n)^2}{\Gamma(2 h_n)} \frac{1}{k_c'(h_n)}~.
\ee
In the large $q$ limit, one can solve $k_c(h_n) = 1$ to derive, 
\be \label{hnq}
h_n = 2 n + 1 + 2 \epsilon_{n}~, \ \ \ \  \epsilon_{n} = \frac{1}{q} \frac{2 n^2 +n + 1}{2 n^2 +n - 1}~, \  \ \ \ \ n\geq1~,\ \ \ q\gg 1~,
\ee
while the OPE coefficients in the large $q$ limit behave as,
\be \label{cnLarge}
c_n^2 = \eps_n^2 \frac{n(1+2n)}{\(n(1+2n)+1\)\(n(1+2n)-1\)}\frac{\sqrt{\pi} \Gamma(2n+1)}{\Gamma(2n+\frac{1}{2}) 2^{4n-2}}~, \ \ \ \ \ q \gg 1~.
\ee
Note that, since the OPE coefficients  $c_n$  vanish for large $q$, the operators $\mO_n$ decouple from the fermions in this limit. Nevertheless, we can study these operators, their correlation functions and dual counterparts, even when $ q= \infty$.

The expression for the four-point function in Eq.~\ref{4pt} is valid for $0<x<1$. 
As a result of anticommutation, 
\be
\langle \chi_i(\tau_1) \chi_i(\tau_2) \chi_j(\tau_3) \chi_j(\tau_4)\rangle= -\langle \chi_i(\tau_2) \chi_i(\tau_1) \chi_j(\tau_3) \chi_j(\tau_4)\rangle~,
\ee
so that  if we are interested in the four-point function for $x<0$, we can exchange $\tau_1 \leftrightarrow \tau_2$, and consider the four-point function with the cross-ratio, 
\be
y = -\frac{\tau_{12} \tau_{34}}{\tau_{23} \tau_{14}}~.
\ee
If $x<0$, then it follows that $y>0$.~\footnote{ The four-point function for $x>1$ is slightly different from (\ref{4pt}), but will not be needed for our purposes. }

\subsubsection*{The dimension-two block}
Finally, the four-point function has a contribution from the dimension-two block, $\mF_{h=2}$, which breaks conformal invariance.  This contribution to the four-point function can be viewed as arising from the Schwarzian action \cite{Kitaev}. In the infrared, SYK  is not truly a CFT$_1$, but just ``nearly'' a CFT$_1$. This is consistent with the bulk being not truly AdS$_2$, but just ``nearly'' AdS$_2$. This must be the case, since AdS$_2$ is two-dimensional and any finite energy excitation has large back-reaction. One cures this by regulating the AdS$_2$, as was understood in \cite{AP}. In particular, AdS$_2$ should be regarded as being embedded in a higher dimensional space, for instance as the near-horizon  limit of an extremal black hole. The ambient geometry serves as the regulator. One can have an effective description of the ``nearly'' AdS$_2$ part if one introduces a dilaton, where the dilaton sets the size of the compact extra dimensions. If the dilaton profile were constant, then one would have pure AdS$_2$. However, the dilaton is not constant and has a nontrivial action, of the type studied in \cite{AP}. It was recognized in   \cite{MSY, Jensen:2016pah} (see also \cite{Engelsoy:2016xyb, Almheiri:2016fws, Bagrets:2016cdf, Mandal:2017thl,Cvetic:2016eiv}) that this action is in fact the same as the Schwarzian action describing the dimension-two operator of SYK. This, of course, should be the case. We will not discuss the $\mF_{h=2}$ block further, as our interest is in the purely conformal sector of SYK: the higher dimension operators and correspondingly the interactions of the  dual bulk fields amongst themselves.

\subsection{The operator product expansion of the fermions} \label{sec:OPE}
The four-point function is written in Eq.~\ref{4pt} as a sum over conformal blocks of operators of dimensions $h_n$. In the limit in which  $\tau_1$ approaches $\tau_2$, the hypergeometric function can be replaced by $1$, giving
\be  \label{F12}
\mF_c(\tau_1, \tau_2, \tau_3, \tau_4) =  G(\tau_{12}) G(\tau_{34})\sum c_n^2\, \Big|\frac{\tau_{12} \tau_{34}}{\tau_{23}\tau_{24}}\Big|^{h_n}~, \ \ \ \ \ |\tau_{12}|\ll 1~.
\ee 
If, in addition, $\tau_3$ approaches $\tau_4$,
\be \label{4ptLim}
\mF_c(\tau_1, \tau_2, \tau_3, \tau_4) = G(\tau_{12}) G(\tau_{34})\,\sum_n c_n^2 \frac{|\tau_{12}|^{h_n}| \tau_{34}|^{h_n}}{|\tau_{24}|^{2 h_n}}~, \ \ \ \ \ |\tau_{12}|\ll 1, \ \ |\tau_{34}|\ll 1~,
\ee
which, as should be the case, is the sum of two-point functions of operators of dimensions $h_n$ that appear in the OPE of the fermions.
These primary, $O(N)$ invariant,  bilinear, operators of dimension $h_n$ can be seen to be, 
\be \label{O2n1}
\mO_{n} =\frac{1}{\sqrt{N}} \sum_{i=1}^N \sum_{k=0}^{2n+1} d_{n k}\, \partial_{\tau}^{k} \chi_i\, \partial_{\tau}^{2n+1 - k} \chi_i~,
\ee
where the coefficients $d_{n k}$ are constructed so as to ensure that the operators are primary, see Appendix~\ref{ap:free}. At weak coupling the dimensions of $\mO_n$ are just  $2n+1$. The dimensions at strong coupling are shifted by an order-one amount, but for large $q$ the shift decays as $1/q$, see Eq.~\ref{hnq}.

The OPE is then given by, 
\be \label{OPE}
\frac{1}{N} \sum_{i=1}^N \chi_i(\tau_1) \chi_i (\tau_2) = \frac{1}{\sqrt{N}}\sum_{n=1}^{\infty} c_n\, \mC_n(\tau_{12}, \partial_2) \mO_n(\tau_2)~,
\ee
where,
\be \label{mCn}
\mC_n(\tau_{12}, \partial_2) = G(\tau_{12}) |\tau_{12}|^{h_n} \( 1+ \frac{1}{2} \tau_{12} \partial_{2} + \ldots\)~.
\ee
In writing (\ref{OPE}) we have left out the identity operator and the $h=2$ operator. Using the OPE in the short time limit, when $\tau_1 \to \tau_2$ and $\tau_3 \to \tau_4$, in the four-point function, we have,
\be \label{Fcnm}
\mF_c= \sum_{n, m} c_n c_m\, \mC_n(\tau_{12},\partial_2) \mC_m(\tau_{34},\partial_4) \langle \mO_n(\tau_2) \mO_m(\tau_4)\rangle~.
\ee
Comparing with (\ref{4ptLim}), we see that, with our conventions, the two-point function of the  bilinears comes with the normalization, 
\be  \label{OO}
\langle\mO_n(\tau_2) \mO_m(\tau_4)\rangle = \frac{\delta_{n,m}}{|\tau_{2 4}|^{2 h_n}}~.
\ee
If one were to write out the full series for (\ref{mCn}) and sum it, then (\ref{Fcnm}) would reproduce the hypergeometric functions in the four-point function (\ref{4pt}).~\footnote{An alternative way to find the four-point function would be to directly find $c_n$ by computing $\langle \chi_i(\tau_1) \chi_i(\tau_2) \mO_n(\tau_0)\rangle $. Thinking of $\mO_n$ as a composite of two fermions, the equation satisfied by this three-point function is the CFT analogue of the Bethe-Salpeter equation;  instead of determining the masses of the bound states of the fermions, it determines the dimensions $h_n$ of $\mO_n$. See \cite{GR}. }

\subsection{Particles in the bulk} \label{sec:bulkpre}
Let us summarize what we know so far about the bulk dual of SYK. The fermion four-point function revealed a tower of $O(N)$ invariant, bilinear, primary operators with dimensions $h_n$.
The standard AdS/CFT dictionary maps such $O(N)$ invariant CFT operators to massive fields, $\phi_n$, in the AdS bulk, as described by,
\be \label{Lbulk0}
\int d^2 x\sqrt{g}\sum_n\[ \frac{1}{2} (\partial\phi_n)^2 + \frac{1}{2} m_n^2 \phi_n^2\]~,
\ee
where the masses of the bulk fields are related  to the dimensions of $\mO_n$ by, $m_n^2 = h_n(h_n-1)$, in units where the radius of the dual AdS space is one.  At leading order in $1/\sqrt{N}$ these fields are free. At next order, they are  expected to have a cubic coupling, of order $1/\sqrt{N}$, whose value is fixed by the six-point function of the fermions. We now turn to determining this cubic coupling.

\section{Three-Point Function of Bilinears} \label{sec:6pt}
In this section we compute the fermion six-point function, which constitutes the main technical result of the paper. There are two classes of Feynman diagrams that contribute at leading nontrivial order in $1/N$: the ``contact'' diagrams  and the planar diagrams. We study these two classes separately, as they are different both technically and conceptually. We let $\mS_1$ denote the sum of all the ``contact'' diagrams, and let $\mS_2$ denote the sum of the planar diagrams. In Sec.~\ref{sec:six1} we write down expressions for $\mS_1$ and $\mS_2$ as integrals of a product of three four-point functions. 

 In Sec.~\ref{sec:OPElim} we discuss how to extract what we are truly interested in, the three-point functions of the  bilinears $\mO_n$, from the fermion six-point function. Roughly, bringing two fermions together gives a superposition of bilinears, so one should take the triple short time limit of the fermion six-point function, which is then proportional to  three-point functions of the bilinear operators.  These bilinear three-point functions have a functional form that is completely fixed by conformal invariance, up to a constant factor, $c_{n m k}$. This coefficient can be written as a sum of two terms, $c_{n m k} = c_{n m k}^{(1)} + c_{n m k}^{(2)}$, with the first arising from the sum of ``contact'' diagrams and the second from the planar diagrams. We will write down an expression for $c_{n m k}^{(1)}$ in terms of a three-loop integral $I_{n m k}^{(1)}$, and an expression for $c_{ n m k}^{(2)}$ in terms of a four-loop integral $I_{n m k}^{(2)}$. All of these results are for general $q$.
  
 The final step is to evaluate the integrals $I_{n m k}^{(1)}$ and $I_{n m k}^{(2)}$. 
 In Sec.~\ref{sec:evInt} we introduce a simple method that allows us to evaluate the integrals $I_{n m k}^{(1)}$ and $I_{n m k }^{(2)}$ at large $q$. The result will be that $c_{n m k}^{(1)}$ takes a simple form, just a product of several Gamma functions, while $c_{n m k}^{(2)}$ will be simply related to the three-point function of bilinears in a generalized free field theory. The latter is computed in Appendix.~\ref{ap:free}. Both $c_{n m k}^{(1)}$ and $c_{n m k}^{(2)}$ are independent of $q$ in the large $q$ limit. 
 
\subsection{Fermion six-point function} \label{sec:six1}
The six-point function of the fermions, to first nontrivial order in $1/N$, can be written as,\vspace{-.37cm}
\begin{multline} \label{Six} 
\frac{1}{N^3} \sum_{i, j, l = 1}^N \langle \chi_i (\tau_1) \chi_i (\tau_2) \chi_j (\tau_3) \chi_j (\tau_4) \chi_l (\tau_5) \chi_l (\tau_6)\rangle = G(\tau_{12}) G(\tau_{34}) G(\tau_{56}) \\
\!\!\!+ \frac{1}{N} \Big( G(\tau_{12}) \mF(\tau_3, \tau_4, \tau_5, \tau_6) 
+   G(\tau_{34}) \mF(\tau_1, \tau_2, \tau_5, \tau_6)+  G(\tau_{56}) \mF(\tau_1, \tau_2, \tau_3, \tau_4)\Big) + \frac{1}{N^2} \mS(\tau_1, \ldots, \tau_6)~.
\end{multline}
The first piece is completely disconnected. The second is partially disconnected and involves the four-point function $\mF$.~\footnote{To be consistent, the $\mF$ appearing in the $1/N$ piece should include not just the ladder diagrams, but also the $1/N^2$ contributions to the four-point function.} 
 The last piece,  $\mS$, is the interesting one. There are two classes of diagrams contributing to $\mS$: the ``contact'' diagrams, whose sum we denote by $\mS_1$, and the planar diagrams, whose sum we denote by $\mS_2$, 
\be \label{SS12}
\mS = \mS_1 + \mS_2~.
\ee

\begin{figure}[t]
\centering
\subfloat[]{
\includegraphics[width=6.5in]{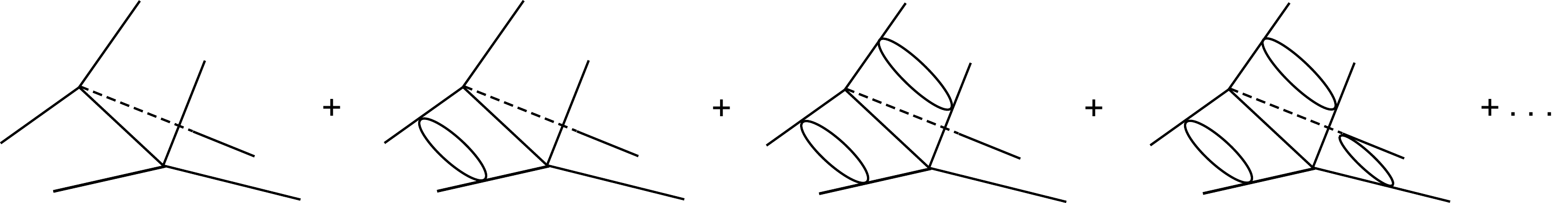}
}\\
\subfloat[]{
\includegraphics[width=2.4in]{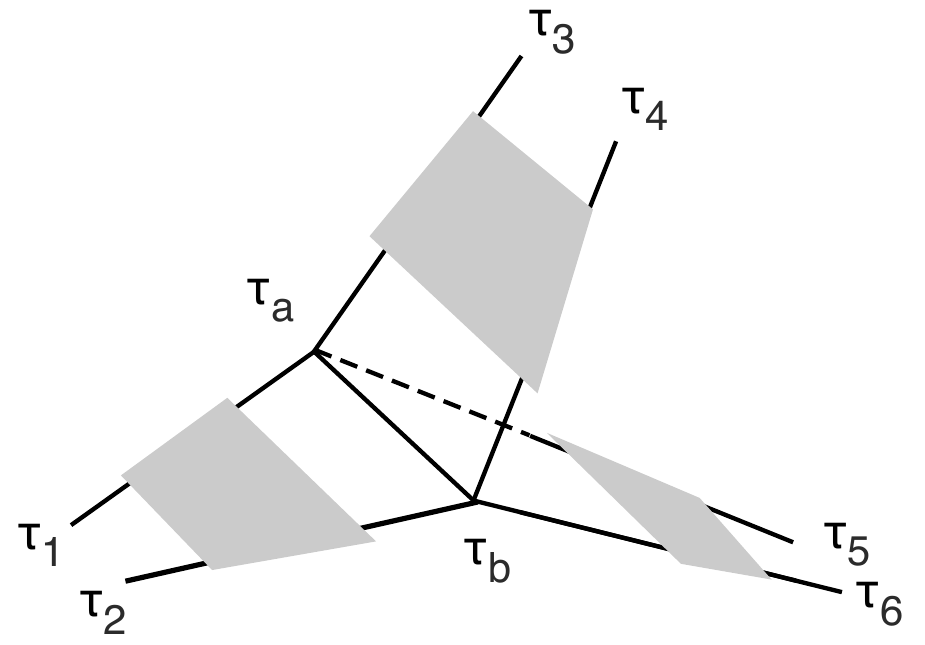}
}
\caption{ The first set of diagrams (``contact'' diagrams) contributing to the six-point function at order $1/N^2$. (a). These diagrams  can be compactly drawn as three four-point functions (shaded squares) glued together as in (b).} \label{Fig6pt0}
\end{figure}

We have illustrated a few of the diagrams contributing to $\mS_1$ in Fig.~\ref{Fig6pt0} (a). These diagrams are simply three four-point functions glued to two interaction vertices connected by $q-3$ propagators, as shown in Fig.~\ref{Fig6pt0} (b),~\footnote{
The combinatorial factor out front in Eq.~\ref{S1} is due to the following contributions: a $1/(q!)^2$ from the square of the coefficient in front of the interaction in the Lagrangian (\ref{SYK}), a factor of $q (q-1) (q-2)$ from contracting the three lines at $\tau_a$ with lines from $\tau_1, \tau_5, \tau_6$, the same factor from the analogous thing at $\tau_b$, a $(q-3)!$ from the ways of contracting the remaining $q-3$ lines going from $\tau_a$ to $\tau_b$, and finally a factor of $(q-1)!$ from the disorder average (\ref{disA}).}
\be \label{S1}
\mS_1 = (q-1)(q-2) J^2 \int d\tau_a d\tau_b\, G(\tau_{ab})^{q-3} \mF(\tau_1, \tau_2, \tau_a, \tau_b) \mF(\tau_3, \tau_4, \tau_a, \tau_b)\mF(\tau_5, \tau_6, \tau_a, \tau_b)~.
\ee

\begin{figure}[t]
\centering
\subfloat[]{
\includegraphics[width=6in]{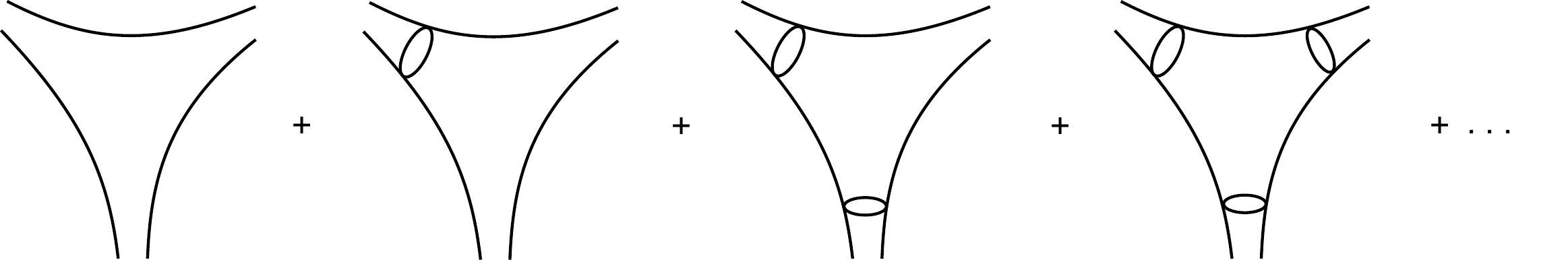}}\\
\subfloat[]{\includegraphics[width=2.3in]{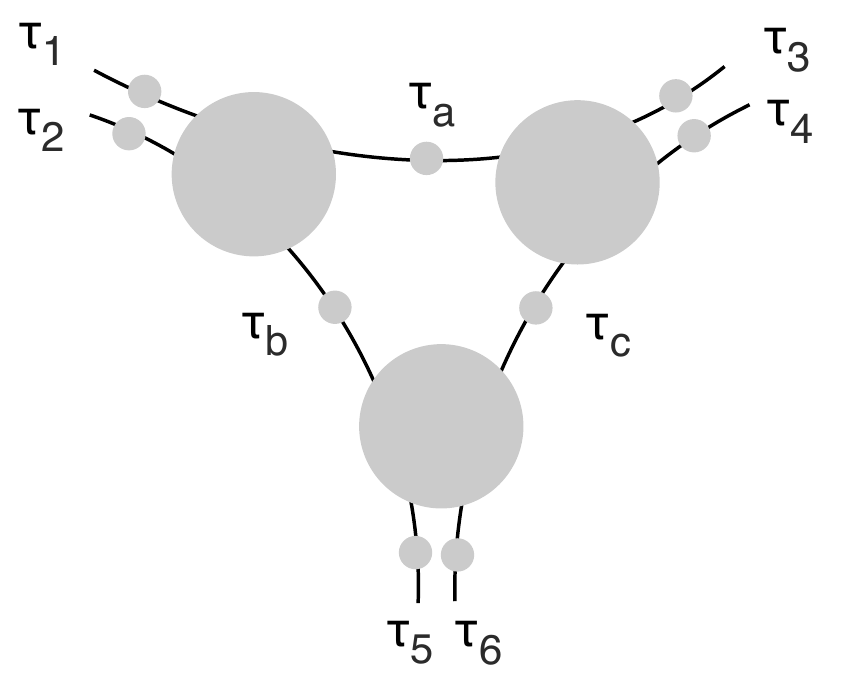}
}
\caption{The second set of diagrams  (planar diagrams)  contributing to the six-point function at order $1/N^2$. These diagrams  can be compactly drawn as three (partially amputated) four-point functions glued together, see Eq.~\ref{6pt1}. The large shaded circles are the 1PI four-point functions. The small shaded circles denote dressing the two-point function; we have made this explicit here, while it is implicit in all other figures.} \label{Fig6pt}
\end{figure}

The other term in the six-point function, $\mS_2$, is given by a sum of  planar diagrams  such as the ones shown in Fig.~\ref{Fig6pt}.  This term is also essentially three four-point functions glued together.  However, to avoid double counting, what we  glue together is   three partially amputated  $\mF$'s, in which the propagator on one of the legs is removed. Defining $\mFa(\tau_1, \tau_2, \tau_3, \tau_4) $ to be $\mF(\tau_1, \tau_2, \tau_3, \tau_4)$ with the outgoing propagator on the $\tau_4$ leg removed, then,~\footnote{This expression is valid at leading order in $1/N$, but not beyond that.}
\be \label{6pt1}
\mS_2 = \int d\tau_a d\tau_b d\tau_c\, \mFa(\tau_1, \tau_2, \tau_a, \tau_b) \mFa(\tau_4, \tau_3, \tau_c, \tau_a) \mFa(\tau_5, \tau_6, \tau_b, \tau_c) ~,
\ee
where the partially amputated $\mF$ is, 
\be
\mFa(\tau_1, \tau_2, \tau_3, \tau_4) = -\int d\tau_0\, \mF(\tau_1, \tau_2, \tau_3, \tau_0)  \int \frac{d\omega_4}{2\pi} e^{- i\omega_4 \tau_{40}} \frac{1}{G(\omega_4)}~.
\ee
With the conformal propagator (\ref{Gstrong}), $\mFa$ becomes,~\footnote{A simple way to see this is to use the Schwinger-Dyson equations for the propagator and self-energy in the infrared, $-G(\omega)^{-1} = \Sigma(\omega)$ and $\Sigma(\tau) = J^2 G(\tau)^{q-1}$.} 
\be \label{mfaE2}
\mFa(\tau_1, \tau_2, \tau_3, \tau_4) =  J^2 \int d\tau_0\, \mF(\tau_1, \tau_2, \tau_3, \tau_0)\, G(\tau_{40})^{q-1}~.
\ee 

\subsection{The short time limit} \label{sec:OPElim}
Having written the fermion six-point function, we can now extract the three-point function of the bilinear operators $\mO_n$. By conformal invariance, this correlation function must have the form,
\be \label{OOO}
\langle \mO_{n}(\tau_2) \mO_{m}(\tau_4) \mO_{k}(\tau_6)\rangle =  \frac{1}{\sqrt{N}}\frac{ c_{n m k}}{|\tau_{24}|^{h_{n} +h_{m}  - h_{k}}| \tau_{46}|^{h_{m} + h_{k} - h_{n}} |\tau_{62}|^{h_{k} + h_{n} - h_{m}}}~.
\ee
In the six-point function of the fermions  we take the short time limit:  $\tau_1 \to \tau_2,\, \tau_3 \to \tau_4,\,\tau_5 \to \tau_6$ of the fermions,  and use for each limit the operator product expansion (\ref{OPE}), keeping the leading term for $|\tau_{12}|, |\tau_{34}|, |\tau_{56}| \ll 1$. This gives, 
\be \label{3pt}
\mS \rightarrow N^{\frac{1}{2}} G(\tau_{12}) G(\tau_{34}) G(\tau_{56}) \sum_{n, m, k} c_n c_m c_k \, |\tau_{12}|^{h_n } |\tau_{34}|^{h_m} |\tau_{56}|^{ h_k} \langle \mO_{n}(\tau_2) \mO_{m}(\tau_4) \mO_{k}(\tau_6)\rangle~,
\ee
showing how $c_{n m k}$ can be extracted from  the six-point function in this triple short time limit. 

Consider first the contribution of the ``contact'' diagrams to the six-point function, $\mS_1$ (\ref{S1}). In the limit  $|\tau_{12}|, |\tau_{34}|, |\tau_{56}| \ll 1$, and making use of the expression for the four-point function in this limit (\ref{F12}), we get
\be \label{S1limit}
\mS_1 \rightarrow b^{q} (q-1)(q-2) G(\tau_{12}) G(\tau_{34}) G(\tau_{56})\sum_{n, m, k} c_n^2 c_m^2 c_k^2 |\tau_{12}|^{h_n}|\tau_{34}|^{h_m}|\tau_{56}|^{h_k} I_{n m k}^{(1)}(\tau_2, \tau_4, \tau_6)~,
\ee
where,
\be \label{I1}
I_{n m k}^{(1)}(\tau_1, \tau_2, \tau_3) = \int d\tau_a d\tau_b \frac{|\tau_{ab}|^{h_n+h_m + h_k-2}}{|\tau_{1a}|^{h_n}|\tau_{1 b}|^{h_n}|\tau_{2a}|^{h_m}|\tau_{2 b}|^{h_m}|\tau_{3 a}|^{h_k} |\tau_{3 b}|^{h_k}}~.
\ee

Now consider the contribution of the planar diagrams, $\mS_2$ (\ref{6pt1}), in the same limit. For this we need the partially amputated four-point function, $\mFa(\tau_1, \tau_2, \tau_3, \tau_4)$ (\ref{mfaE2}), in the limit $|\tau_{34}|\ll 1$. Again using (\ref{F12}) and evaluating the $\tau_0$ integral with the help of the formula (\ref{cf3ptint}) in Appendix~\ref{ap:integrals}, 
\be \label{Famp2}
\!\!\!\!\! \mFa(\tau_1, \tau_2, \tau_3, \tau_4) = G(\tau_{12})  \sum_n c_n^2\, \xi_n\, \sgn(\tau_{23} \tau_{24}) \frac{|\tau_{12}|^{h_n} |\tau_{34}|^{h_n-1}}{|\tau_{23}|^{h_n-1 + 2\Delta} |\tau_{24}|^{h_n+1 - 2\Delta}}~, \, \, \, \, \,  |\tau_{34}|\ll 1~,
\ee
where, 
\be \label{eq:xi}
\xi_n =  b^q\, \pi^{\frac{1}{2}} \frac{\Gamma(1 - \Delta + \frac{h_n}{2})}{\Gamma(\frac{1}{2} + \Delta - \frac{h_n}{2})}\frac{\Gamma(\frac{1}{2} - \frac{h_n}{2})}{\Gamma(\frac{h_n}{2})}   \frac{\Gamma\(\Delta\)}{\Gamma\( \frac{3}{2} - \Delta\)}~.
\ee
In the large $q$ limit $\xi_n$ simplifies to, 
\be
\xi_n = \(n + \frac{1}{2}\) \( 1- \frac{1}{\eps_n q}\) = {{ 2n+1}\over {2n^2+2n+1}}  ~, \ \ \ \ \ q  \gg1~.
\ee
Inserting $ \mFa$ into the expression for $\mS_2$ (\ref{6pt1}), we find that, in the limit  $|\tau_{12}|, |\tau_{34}|, |\tau_{56}| \ll 1$, 
\be\label{S2limit}
\mS_2 \rightarrow G(\tau_{12})G(\tau_{34})G(\tau_{56}) \sum_{n, m, k} c_n^2 \xi_n c_m^2 \xi_m c_k^2 \xi_k\, |\tau_{12}|^{h_n} |\tau_{34}|^{h_m} |\tau_{56}|^{h_k}\, I_{n m k}^{(2)} (\tau_2, \tau_4, \tau_6)~,
\ee
where, 
\be \label{I2}
\!\!\!\!\!\!\!\!\! I_{n m k}^{(2)}(\tau_1, \!\tau_2, \!\tau_3) \!= \!\!\! \int\!\! d\tau_a d\tau_b d\tau_c \frac{  -\sgn(\tau_{1a} \tau_{1b} \tau_{2a} \tau_{2c} \tau_{3b} \tau_{3c}) \! |\tau_{ab}|^{h_n  \!- \!1} |\tau_{ca}|^{h_m \!- \!1} |\tau_{b c}|^{h_k  \!- \!1}}{|\tau_{1a}|^{h_n \!- \!1+2\Delta} |\tau_{1b}|^{h_n +1-2\Delta} |\tau_{2 c}|^{h_m  \!- 1\!+2\Delta} |\tau_{2a}|^{h_m+1-2\Delta} |\tau_{3 b}|^{h_k  \!-1 \!+2\Delta} |\tau_{3 c}|^{h_k +1-2\Delta}}.
\ee
We will evaluate  the integrals $I_{n m k}^{(1)}$ (\ref{I1}) and  $I_{n m k}^{(2)}$ (\ref{I2}) in the next section. Here we note that they both transform as conformal three-point functions, 
\be \label{6intA}
I_{n m k}^{(\nu)} (\tau_1, \tau_2, \tau_3) =\frac{\mI^{(\nu)}_{n m  k } }{ |\tau_{12}|^{h_n+h_m-h_k}  |\tau_{13}|^{h_n +h_k - h_m} |\tau_{23}|^{h_m+h_k -h_n}}~,
\ee
where $\nu = 1, 2$.  This can be verified by showing that (\ref{I1}), (\ref{I2}) and (\ref{6intA}) transform in the same way under $SL(2, \mathcal{R})$ transformations: $ \tau \to  {a\tau+b \over c\tau +d}$. We can thus write the coefficient $c_{n m k}$ of the three-point function  of the bilinears (\ref{OOO}) as, 
\be\label{cnmk}
c_{n m k} = c_{n m k}^{(1)} + c_{ n m k}^{(2)}~, 
\ee
where $c_{n m k}^{(1)}$ follows by comparing (\ref{S1limit})  and (\ref{3pt}), 
\be \label{cnmk1}
c_{n m k}^{(1)} = c_n c_m c_k\,  b^q (q-1)(q-2)\, \mI_{n m k}^{(1)}~,
\ee
and $c_{n m k}^{(2)}$ follows by comparing (\ref{S2limit}) and (\ref{3pt}), 
\be \label{cnmk2}
c_{ n m k}^{(2)} = c_n c_m c_k\,  \xi_n \xi_m \xi_k\, \mI_{n m k}^{(2)}~.
\ee

\subsection{Evaluating the integrals}  \label{sec:evInt}
\begin{figure}[t]
\centering
\subfloat[]{
\includegraphics[width=1.8in]{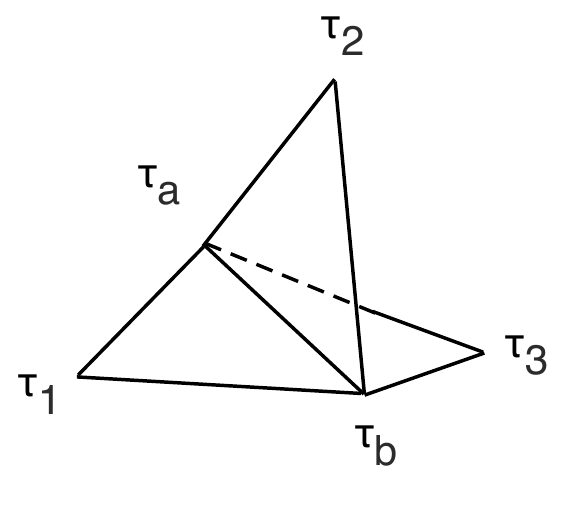}} \ \ \ \ \ \  \ \ \ \ \ \ \ \ \
\subfloat[]{\includegraphics[width=2.0in]{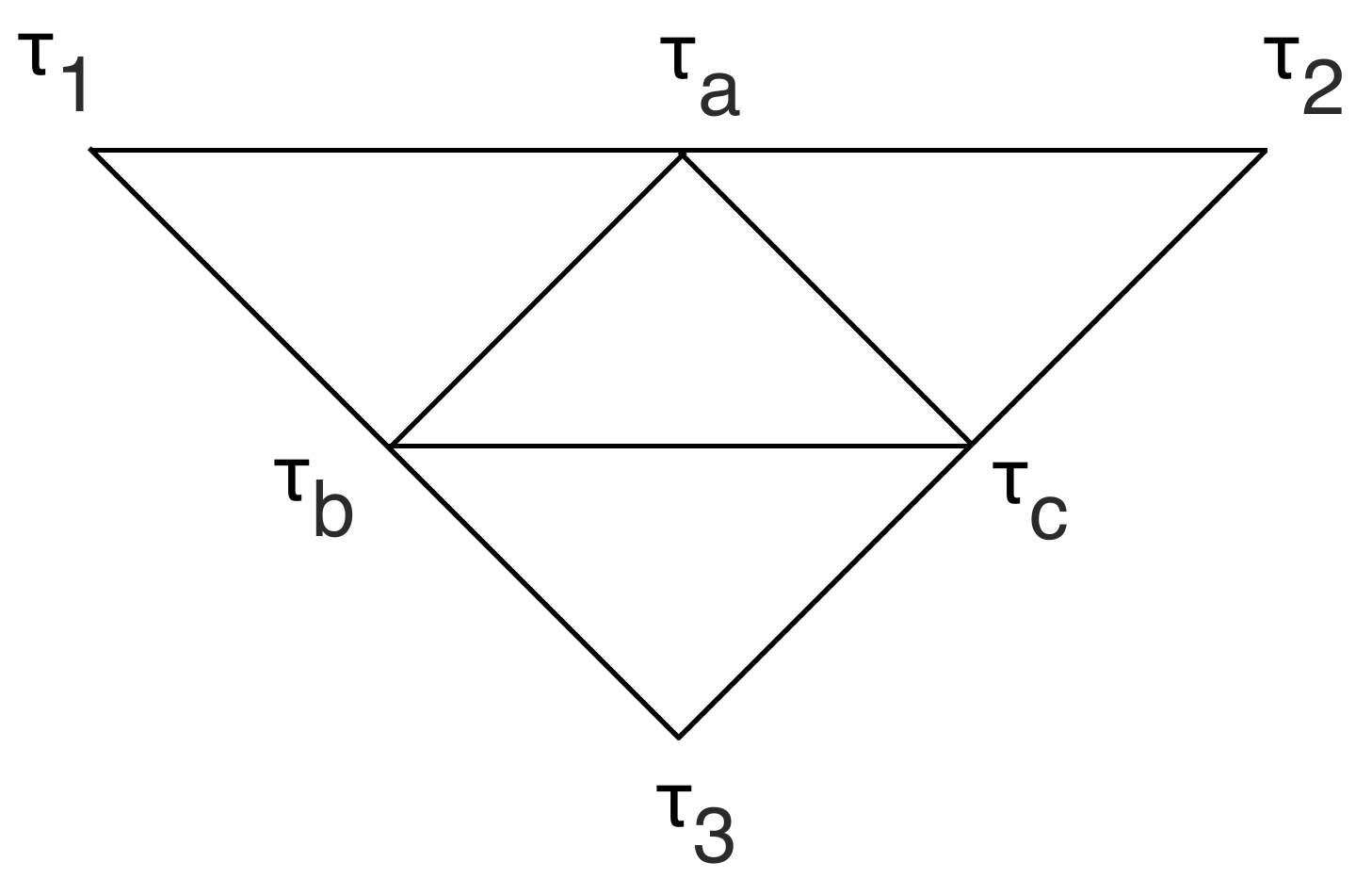}}
\caption{The integrals $I_{n m k}^{(1)}$ (\ref{I1}) and $I_{n m k}^{(2)}$ (\ref{I2}),  relevant for the three-point function of the fermion bilinears, can be viewed as  three-loop and four-loop integrals, respectively.} \label{Fig3loop}
\end{figure}
Having found the coefficient $c_{n m k}$ of the three-point function of the bilinears expressed in terms of the coefficients $\mI_{nmk}^{(1)}$ and  $\mI_{nmk}^{(2)}$ of the integrals $I_{n m k}^{(1)}$ (\ref{I1}) and $I_{n m k}^{(2)}$ (\ref{I2}), respectively, all that is left to do is to evaluate these integrals. They can be regarded as three and four-loop integrals, as shown in Fig.~\ref{Fig3loop}.

For any fixed $n, m, k$, one can evaluate $I_{n m k}^{(1)}$ and $I_{n m k}^{(2)}$ using standard Mellin-Barnes techniques \cite{Smirnov, Henn}. However, there is little reason to expect the answer to be simple. Indeed, for general $q$-body SYK, the dimensions $h_n$ of the bilinears do not even have an  explicit form, being determined implicitly through the solution of $k_c(h)=1$, with $k_c(h)$  defined in (\ref{kch}). However, there is considerable simplification in the limit  of large $q$, as was mentioned before in our discussions of the fermion two-point and four-point functions. One should note that, although SYK simplifies in the large $q$ limit, it is certainly far from trivial in this limit.
 
 The simplification of the bilinear three-point function occurs at large $q$ because then the IR fermion dimension $\Delta= 1/q$ is small, and the dimensions of the bilinears are close to odd integers, $h_n = 2n + 1 + 2 \eps_n$. In this section we evaluate $I_{n m k}^{(1)}$ and $I_{n m k}^{(2)}$ to leading order in the large $q$ limit. 

The   conformal integrals we will need to consider are somewhat similar to integrals encountered in the study of four-dimensional amplitudes, for example in  \cite{Smirnov:1999gc, Drummond:2006rz}. In that case, one is actually in $4-\ep$ dimensions: the small $\ep$ is analogous to our small $1/q$. A technique for evaluating these integrals is to employ the Melin-Barnes representation to transform the original integral into integrals of products of Gamma functions. One then reorganizes the expression,  finding the poles which give the most divergent contribution as $\ep$ goes to zero. 

We will not use this technique. Rather, since we are only interested in the leading behavior at large $q$, it is faster to pick out the most divergent terms at the outset, immediately turning an integration problem into an algebraic problem.

\subsubsection{$I_{n m k}^{(1)}$ }
\vspace{.2cm}
\noindent \textit{$I_{n m k}^{(1)}$ at large $q$}
\vspace{.2cm}

We start with the integral $I_{n m k}^{(1)}$ (\ref{I1}). This integral can actually be evaluated exactly, but we first warm up with  large $q$ limit. First, we note that the integral is convergent in the IR: in the limit of  large $\tau_a$, the integrand decays as $1/\tau_a^2$. Similarly for $\tau_b$.  However, there are UV divergences potentially occurring in nine different regions selected from: $\tau_a, \tau_b \rightarrow \tau_1, \tau_2, \tau_3$, where the integrand blows up. The divergence is algebraic, $\sim \({1\over 0}\)^{h_n} $, and consequently, in the limit of infinite $q$, for which $h_n \rightarrow 2n+1$, the integral over $\tau_a$ will diverge as $q$, as $q\to \infty$. Similarly the integral over  $\tau_b$ will yield a factor of 
$q$, as $q	 \to \infty$. These will cancel with the factors of $q$ in (\ref{cnmk1}), yielding a finite result for $c_{n m k}^{(1)}$.

Consider first the integrand in the region $\tau_a\rightarrow \tau_1,\, \tau_b\rightarrow \tau_2$, where we can write,
\be \label{I11}
I_{n m k}^{(1)} = \ldots + \int d\tau_a d\tau_b \frac{\sgn(\tau_{ab} \tau_{1a} \tau_{2 a} \tau_{3a} \tau_{1 b} \tau_{2 b} \tau_{3b})}{|\tau_{1a}|^{2 \eps_n}|\tau_{2b}|^{2 \eps_m}} f_{n m k}^{(1)} + \ldots
\ee
where 
\be \label{f}
f_{n m k}^{(1)} = \frac{\tau_{ab}^{2n +2m +2 k+1}}{\tau_{1a}^{2n+1} \tau_{1b}^{2n+1} \tau_{2a}^{2m+1} \tau_{2 b}^{2m+1} \tau_{3a}^{2k+1}\tau_{3b}^{2k+1}}~.
\ee
We have used the fact that, since $|\tau_{1a}|, |\tau_{2 b}| \ll 1$, we are justified in dropping the $\eps_n$ occurring in most of the terms in the integrand of $I_{n m k}^{(1)}$. In other words, a term like $\tau_{2a}^{2m+1+ 2\eps_n}$ is equal to $\tau_{ 2a}^{2m+1}$, to leading order in $1/q$, as long as $\tau_1$ is not close to $\tau_2$.  As a result, the function $f_{n m k}^{(1)}$ is holomorphic,  allowing us to do a series expansion in powers of $\tau_{1a}$ and $\tau_{2b}$, 
\be \label{fseries}
f_{n m k}^{(1)} = \ldots + \frac{1}{\tau_{1 a} \tau_{2b}} \frac{-s_{n m k}^{(1)}}{\tau_{12}^{2n+2m-2k+1}\tau_{13}^{2n+2k-2m+1}\tau_{23}^{2m+2k -2n+1}} + \ldots~,
\ee
where we have singled out the term we are interested in, the one that scales as $1/(\tau_{1 a} \tau_{2 b})$. This expansion  confirms that the integral is a conformal three-point function, as above, with  $s_{ n m k}^{(1)}$ denoting the coefficient of the relevant term.  The contribution to the integral from this piece of the integrand is then,
\be \nonumber
\!\!\!\! I_{n m k}^{(1)} = \ldots +\frac{s_{n m k}^{(1)}}{|\tau_{12}|^{2n+2m-2k+1}|\tau_{13}|^{2n+2k-2m+1}|\tau_{23}|^{2m+2k -2n+1}} \int_{\tau_a \approx \tau_1}\!\!\! \!d\tau_a\frac{\!\!\!\!1}{\, |\tau_{1a}|^{1+2 \eps_n}} \int_{\tau_b \approx \tau_2}\!\!\! \! d\tau_b \frac{\!\!\!\!1}{\, |\tau_{2b}|^{1+2 \eps_m}}+ \ldots
\ee
where,  
\be \label{trivial}
\int_{\tau_a \approx \tau_1} d\tau_a \frac{1}{\,\, |\tau_{1a}|^{1+2 \eps_n}}  = \frac{1}{\eps_n}~.
\ee
There are eight more regions to consider. The terms when $\tau_a \rightarrow \tau_1, \tau_b\rightarrow \tau_3$ or  $\tau_a \rightarrow \tau_2, \tau_b\rightarrow \tau_3$ follow by symmetry (exchanging $\tau_a$ with $\tau_b$ also follows by symmetry). The remaining three regions are of the kind, $\tau_a, \tau_b \rightarrow \tau_1$. However,  expanding $f_{n m k}^{(1)}$ shows that there is no term that scales as $1/(\tau_{1a}\tau_{1b})$.~\footnote{Indeed, there can't possibly be such a term. With $\tau_a, \tau_b \rightarrow \tau_1$, there is no way to produce a power of $\tau_{23}$, which is necessary for the result to take the form of a conformal three-point function.} Thus, we have the result for $I_{n m k}^{(1)}$, at large $q$, is (\ref{6intA}) with, 
\be \label{C1}
\mI_{n m k}^{(1)} = 2 s_{n m k }^{(1)} \frac{\eps_n+ \eps_m + \eps_k}{\eps_n \eps_m \eps_k}~,
\ee
where the expression for $s_{n m k}^{(1)}$ turns out to take a relatively simple form, 
\be \label{C1s}
s_{n m k }^{(1)} =  (-4)^{n + m + k} \frac{\Gamma(\frac{1}{2}\!+k + m\! -\! n) \Gamma(\frac{1}{2}\!+k\! -\!m+n) \Gamma(\frac{1}{2}\!-\! k+m+n) \Gamma(1\!+k\!+m+n)}{\pi^{\frac{3}{2}} \Gamma(1+2n ) \Gamma(1+2m)\Gamma(1+2k)}  ~.
\ee

\vspace{.4cm}
\noindent \textit{$I_{n m k}^{(1)}$ at finite $q$}
\vspace{.2cm}

The result we found for $I_{n m k}^{(1)}$ at large $q$ is sufficiently simple that one may suspect the integral can be evaluated even at finite $q$. Indeed, it can. Starting with  $I_{n m k}^{(1)}$ (\ref{I1}) we do a change of variables $\tau_a \rightarrow \tau_1 - 1/\tau_a$ and $\tau_b \rightarrow \tau_1 - 1/\tau_b$, transforming the integral into,
\be
I_{n m k}^{(1)} = \frac{1}{|\tau_{21}|^{2 h_m} |\tau_{31}|^{2 h_k}} \int d\tau_a d\tau_b \frac{|\tau_{ab}|^{h_n + h_m + h_k - 2}}{|\tau_a + \tau_{21}^{-1}|^{h_m}|\tau_b + \tau_{21}^{-1}|^{h_m} |\tau_{a} + \tau_{31}^{-1}|^{h_k}|\tau_{b} + \tau_{31}^{-1}|^{h_k}}~.
\ee
This is of the form of a generalized Selberg integral, see Appendix~\ref{ap:integrals}. Making use of (\ref{SelFull}), we find,
\be \label{I1exact}
\!\!\mathcal{I}_{n m k}^{(1)}\! =\!\frac{  \sqrt{\pi}\, 2^{h_n + h_m + h_k -1}\, \Gamma(1\!-\!h_n) \Gamma(1\!-\!h_m) \Gamma(1\!-\!h_k)}{\Gamma\(\frac{3 - h_n - h_m - h_k}{2}\)}\! \[\rho(h_n, h_m, h_k)\! +\! \rho(h_m, h_k, h_n)\!+\! \rho(h_k, h_n, h_m) \]~,
\ee
where,
\be
\rho(h_n, h_m, h_k) = \frac{\Gamma(\frac{h_m +h_k - h_n}{2})}{\Gamma(\frac{2-h_n-h_m +h_k}{2})\Gamma(\frac{2-h_n-h_k+h_m}{2})}\( 1+ \frac{\sin(\pi h_m)}{\sin(\pi h_k) - \sin(\pi h_n + \pi h_m)}\)~.
\ee

\subsubsection{$I_{n m k}^{(2)}$ at large $q$} \label{sec:I2largeq}
%\vspace{.2cm}
%\noindent \textit{Method 1}:
%\vspace{.2cm}

We now turn to $I_{n m k}^{(2)}$ (\ref{I2}), using the same method  we used for studying $I_{n m k}^{(1)}$ at large $q$. There are eight regions that can lead to UV divergences: $\tau_a \rightarrow \tau_1, \tau_2$,
$\,\, \tau_b  \rightarrow \tau_1, \tau_3$, $\,\, \tau_c \rightarrow \tau_2, \tau_3$. Let us start with $\tau_a \rightarrow \tau_1,\,\,  \tau_b \rightarrow \tau_3,\, \,  \tau_c\rightarrow \tau_2$,
\be \label{I22}
I_{n m k}^{(2)} = \ldots  - \int d\tau_a d\tau_b d\tau_c \frac{ \sgn(\tau_{1a} \tau_{1b} \tau_{2a} \tau_{2c} \tau_{3b} \tau_{3c}) }{|\tau_{1a}|^{2\eps_n^+} |\tau_{3b}|^{2\eps_k^+}|\tau_{2 c}|^{2\eps_m^+}}\, f_{n m k}^{(2)}+ \ldots,
\ee
where we have defined, 
\be \label{epspm}
\eps_n^+ \equiv \eps_n + \Delta~, \ \ \ \ \eps_n^- = \eps_n - \Delta~, 
\ee
and
\be \label{f20}
 f_{n m k}^{(2)}= \frac{\tau_{ab}^{2n} \tau_{ca}^{2m} \tau_{bc}^{2 k}}{\tau_{1a}^{2n} \tau_{1b}^{2n+2} \tau_{2c}^{2m} \tau_{2a}^{2m+2} \tau_{3b}^{2k} \tau_{3c}^{2k+2}}~.
\ee
We expand $f_{n m k}^{(2)}$ in $\tau_{1a}, \tau_{2c}, \tau_{3b}$, picking out the term $1/(\tau_{1a}\tau_{2c}\tau_{3b})$, 
\be \label{f2}
f_{n m k}^{(2)} = \ldots + \frac{1}{\tau_{1a}\tau_{2c}\tau_{3b}} \frac{-s_{n m k}^{(2)}}{\tau_{12}^{2n+2m-2k+1}\tau_{13}^{2n+2k-2m+1}\tau_{23}^{2m+2k -2n+1}} + \ldots~.
\ee
Inserting this back into the integral (\ref{I22}) and performing the integrals as before in (\ref{trivial}) we get,
\be
\!\!\!\! I_{n m k}^{(2)} = \ldots +\frac{1}{\eps_n^+ \eps_m^+ \eps_k^+}\frac{s_{n m k}^{(2)}}{|\tau_{12}|^{2n+2m-2k+1}|\tau_{13}|^{2n+2k-2m+1}|\tau_{23}|^{2m+2k -2n+1}} + \ldots
\ee
There are seven other regions of $\tau_a, \tau_b, \tau_c$ that lead to UV divergences. In fact, six of them have a slight subtlety: it is important to keep the epsilon's in the numerator if two of the times, for instance $\tau_a$ and $\tau_b$, are approaching the same time. In particular, 
\be
\int_{\tau_a \approx \tau_1}\!\! d\tau_a\, \int_{\tau_b \approx \tau_1}\!\! d\tau_b \, \, \frac{|\tau_{ab}|^{ \eps_1^+ + \eps_2^-}}{|\tau_{a1}|^{1+ 2\eps_1^+}|\tau_{b1}|^{1 + 2\eps_2^-}}  \approx \frac{2}{\eps_1^+ \eps_2^-}~,
\ee
which is twice what one would have gotten if one had dropped the numerator in the integrand. Accounting for all eight regions we get that $I_{n m k}^{(2)}$, at large $q$, is (\ref{6intA}) with, 
\be \label{Cscri2}
\mI_{n m k}^{(2)} = s_{n m k}^{(2)}\( 2 \frac{\eps_n^+ + \eps_m^-}{\eps_n^+ \eps_m^-}\frac{\eps_m^+ + \eps_k^-}{\eps_m^+ \eps_k^-}\frac{\eps_k^+ + \eps_n^-}{\eps_k^+ \eps_n^-} - \frac{1}{\eps_n^+\eps_m^+\eps_k^+} - \frac{1}{\eps_n^- \eps_m^-\eps_k^-}\)~.
\ee
Finding $s_{n m k}^{(2)}$ is straightforward: it simply involves doing a series expansion of $f_{n m k}^{(2)}$ (\ref{f20}). In other words, $s_{n m k}^{(2)}$ is, 
\be \label{smethod1}
s_{n m k}^{(2)} = [\tau_{1a}^{-1}] [\tau_{2c}^{-1}][\tau_{3b}^{-1}]\,  f_{n m k}^{(2)}~,
\ee
where $[x^q]$ denotes the coefficient of $x^q$ of what follows after it.

%Next, we will redo the calculation of the $I_{n m k}^{(2)}$ integral in a slightly different way. We will get the same answer, but in a form that is more useful.

%\vspace{.2cm}
%\noindent \textit{Method 2}:
%\vspace{.2cm}

We would like to be more explicit  as to what $s_{n m k}^{(2)}$ is. To do this, we  rewrite  $I_{n m k}^{(2)}$ as,
\begin{multline} \label{Iv2}
I_{n m k}^{(2)}= -\int d\tau_a d\tau_b d\tau_c\,   (\tau_{a1}+ \tau_{1b})^{2n}  (\tau_{c2}+ \tau_{2a})^{ 2m}  (\tau_{b 3}+\tau_{3 c})^{2 k}\\
\frac{\sgn(\tau_{1a} \tau_{1b} \tau_{2a} \tau_{2c} \tau_{3b} \tau_{3c})\, \, |\tau_{ab}|^{ \eps_n^+ + \eps_n^-}|\tau_{ca}|^{\eps_k^+ + \eps_k^-}\, |\tau_{bc}|^{\eps_m^+ + \eps_m^-}}{|\tau_{1a}|^{2 n + 2\eps_n^+} |\tau_{1b}|^{2 n+ 2\eps_n^- + 2} |\tau_{2 c}|^{2 m+2\eps_m^+ } |\tau_{2a}|^{2 m_{\eps}+ 2\eps_m^-+2} |\tau_{3 b}|^{2 k+2\eps_k^+} |\tau_{3 c}|^{2 k + 2\eps_k^- +2}}~,
\end{multline}
where we have rewritten $\tau_{ab}$ as $\tau_{a1} + \tau_{1b}$, and similarly for $\tau_{ca}$ and $\tau_{bc}$. %The only approximation we made in going from original form to this expression is dropping the $\epsilon$'s appearing in the powers in the numerator.  The justification for dropping  the $\epsilon$'s  is that $\tau_{ab}^{2n+2\eps_n} $ does not contribute a divergent piece to the integral as $\tau_{ab}\to 0$  and so to leading order in $1/q$ it is equivalent to  $\tau_{ab}^{2n} $.
Now, performing a binomial expansion of these terms,
\be
I_{n m k}^{(2)} = \sum_{p_1=0}^{2n}\sum_{p_2=0}^{2m} \sum_{p_3=0}^{2k}    \binom{2n}{p_1}\binom{2m}{p_2}\binom{2k}{p_3} \mathcal{J}_{n, m, k} (\tau_1, \tau_2, \tau_3)~,
\ee
where, 
\begin{multline} \label{329}
\mathcal{J}_{n, m, k} (\tau_1, \tau_2, \tau_3) =  \int d\tau_a d\tau_b d\tau_c \(\sgn(\tau_{a1} \tau_{1b})\)^{p_1+1}(\sgn(\tau_{c2}\tau_{2 a}))^{p_2+1} (\sgn(\tau_{b 3}\tau_{3 c}))^{p_3+1} \\
\frac{|\tau_{ab}|^{ \eps_n^+ + \eps_n^-}  |\tau_{ca}|^{\eps_k^+ + \eps_k^-} |\tau_{bc}|^{\eps_m^+ + \eps_m^-}}{|\tau_{1a}|^{2 n -p_1+ 2\eps_n^+} |\tau_{1b}|^{p_1+2+ 2\eps_n^-} |\tau_{2 c}|^{2 m-p_2+2\eps_m^+ } |\tau_{2a}|^{p_2+2+ 2\eps_m^-} |\tau_{3 b}|^{2 k-p_3 +2\eps_k^+} |\tau_{3 c}|^{p_3+2 + 2\eps_k^- }}
\end{multline}
Next, notice that, 
\be
 \int_{\tau_a \approx \tau_1} d\tau_a \, \frac{1}{|\tau_{a1}|^{q_1+ 2\eps_1}|\tau_{a 2}|^{q_2}} \approx  \frac{1}{\eps_1} \,  \binom{q_1\! +\! q_2\! -\! 2}{q_1\! -\! 1}\, \frac{(\sgn\, \tau_{12})^{q_2}}{(-\tau_{12})^{q_1 + q_2-1}}~,
 \ee
 where we did a binomial expansion to pick out the appropriate term, and applied (\ref{trivial}). In this way, we evaluate (\ref{329}), to find $s_{n m k}^{(2)}$ is the triple sum,
 \be \label{mCSum}
s_{n m k}^{(2)} = -\!\!\!\sum_{p_1, p_2, p_3} \!\binom{2n}{p_1}\!\binom{2m}{p_2}\!\binom{2 k}{p_3}\!
 \binom{2n\!+\!p_2\!-\!p_1}{p_2+1}\!\binom{2m\!+\!p_3\!-\!p_2}{p_3+1}\!\binom{2 k \!+\!p_1\!-\!p_3}{p_1+1}\!
 \frac{z^{p_1-p_2 + 2m - 2 k}}{ (\!-1\!-\!z)^{p_3 - p_2+2n - 2 k}} ~.
 \ee
 where we defined $z = \tau_{12}/\tau_{31}$. 
 This expression is symmetric under all permutations of $n, m, k$.
In addition, this sum must be independent of $z$.  Neither of these properties is manifest, although one can verify that they are both true.   Some properties of this sum are discussed in Appendix~\ref{ap:sum}. In fact, this same sum occurs in the computation of the three-point function of bilinears in a (particular) generalized free field theory, see Appendix~\ref{ap:free}.

\section{The Bulk Cubic Couplings} \label{sec:bulk}
In the previous section we found the coefficients $c_{n m k}$ of the conformal three-point function of the bilinear operators $\mO_n$. In the limit of large $q$ we wrote explicit equations for $c_{n m k}$. In this section we use these $c_{n m k}$ to determine  the  cubic couplings of the bulk fields $\phi_n$ dual to $\mO_n$. 

The bulk Lagrangian, to order $1/\sqrt{N}$, is,
\be \label{Lbulk}
S_{bulk} = \int d^2 x\sqrt{g}\[ \frac{1}{2} (\partial\phi_n)^2 + \frac{1}{2} m_n^2 \phi_n^2 +\frac{1}{\sqrt{N}} \, \lambda_{n m k }\,  \phi_n \phi_m \phi_k\]~.
\ee
One could  also consider cubic terms with derivatives, however, as  shown in Appendix~\ref{Sec:redef}, at this order in $1/N$ they are equivalent to the non-derivative terms up to a field redefinition. 
We use this bulk Lagrangian to compute, via the AdS/CFT dictionary,  the three-point function of the boundary dual. Matching the result with what we found for the SYK three-point function will determine $\lambda_{n m k}$. 

From the  tree level Witten diagram (Fig.~\ref{FigIntroW}),  the three-point function resulting from this bulk Lagrangian is  \cite{Freedman:1998tz},~\footnote{To simplify comparing with the SYK result, we have normalized the operators to have the two-point function, $ \langle O_n (\tau_1) O_m (\tau_2)\rangle =  \delta_{n m}  |\tau_{12}|^{-2 h _n}$.}
\be
\langle O_n (\tau_1) O_m (\tau_2) O_k(\tau_3) \rangle  =\frac{1}{\sqrt{N}}\frac{ \lambda_{n m k } }{N_n N_m N_k} \frac{a_{n m k}}{|\tau_{12}|^{h_n + h_m - h_k} |\tau_{23}|^{h_m +h_k - h_n}|\tau_{13}|^{h_k+h_n - h_m}}~,
\ee
where,
\be \label{anmk}
a_{n m k} = - \frac{\Gamma\[\frac{1}{2} (h_n\!+\! h_m\! -\! h_k)\]\Gamma\[\frac{1}{2} (h_m\! +\! h_k\! -\! h_n)\]\Gamma\[\frac{1}{2} (h_k\! +\! h_n\! -\! h_m)\] \Gamma\[\frac{1}{2}(h_n\! +\! h_m\! +\! h_k\! -\! 1)\]}{2 \pi \Gamma\[ h_n - \frac{1}{2}\]\Gamma\[ h_m - \frac{1}{2}\] \Gamma\[ h_k - \frac{1}{2}\]} ~.
\ee
and 
\be
N_n^2 = \frac{2 h_n - 1}{h_n}\frac{\Gamma(h_n + 1)}{\pi^{\frac{1}{2}} \Gamma\(h_n - \frac{1}{2}\)}~.
\ee

Matching with the three-point function (\ref{OOO}) of the SYK bilinears $\mO_n$, we find that the bulk cubic coefficient $\lambda_{n m k }$ is, 
\be \label{dnmk}
\lambda_{n m k }  = c_{n m k } \frac{N_n N_m N_k}{ a_{n m k}}~.
\ee

When computing $c_{n m k}$ we split it into two terms, $c_{n m k}^{(1)}$ and $c_{n m k}^{(2)}$, resulting from summing the ``contact'' and planar diagrams, respectively. It is useful to similarly split $\lambda_{n m k}$, 
\be
\lambda_{n m k } =\lambda_{n m k}^{(1)} + \lambda_{n m k}^{(2)}~.
\ee
We study each in turn.
\subsection{The ``contact'' diagrams}
\subsubsection{Large $q$}
Combining all the pieces, we find that the ``contact'' diagrams lead to the cubic coupling,
\be \label{lambda1F}
\lambda^{(1)}_{n m k} =  - (-1)^{n + m + k}  \frac{16}{\sqrt{\pi}} q\(\eps_n + \eps_m + \eps_k\)\, \alpha_n \alpha_m \alpha_k~,
\ee
where recall that (\ref{hnq}),
\be
 \epsilon_{n} = \frac{1}{q} \frac{n(2n+1)  + 1}{n(2n+1) - 1}~,
 \ee
 and we have defined, 
\be\label{alphan}
\alpha_n = \sqrt{\frac{ n (1+4n) (1+2n)}{\(n(1+2n) +1\)\(n(1+2n) - 1\)}}~.
\ee
\begin{figure}[t]
\centering 
\subfloat[]{\!\!\!\!\!\!\!
\includegraphics[width=2.2in]{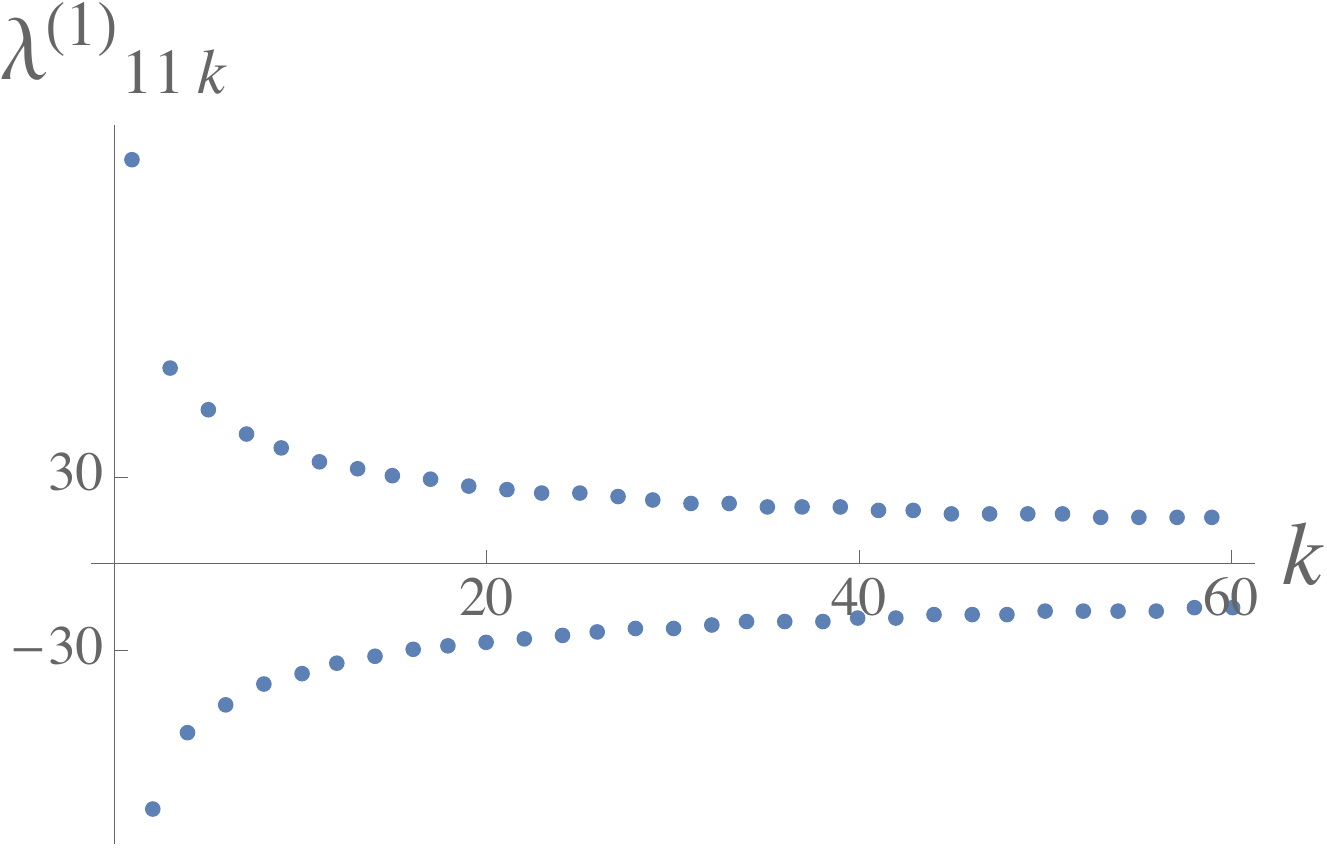}
}
\subfloat[]{
\includegraphics[width=2.1in]{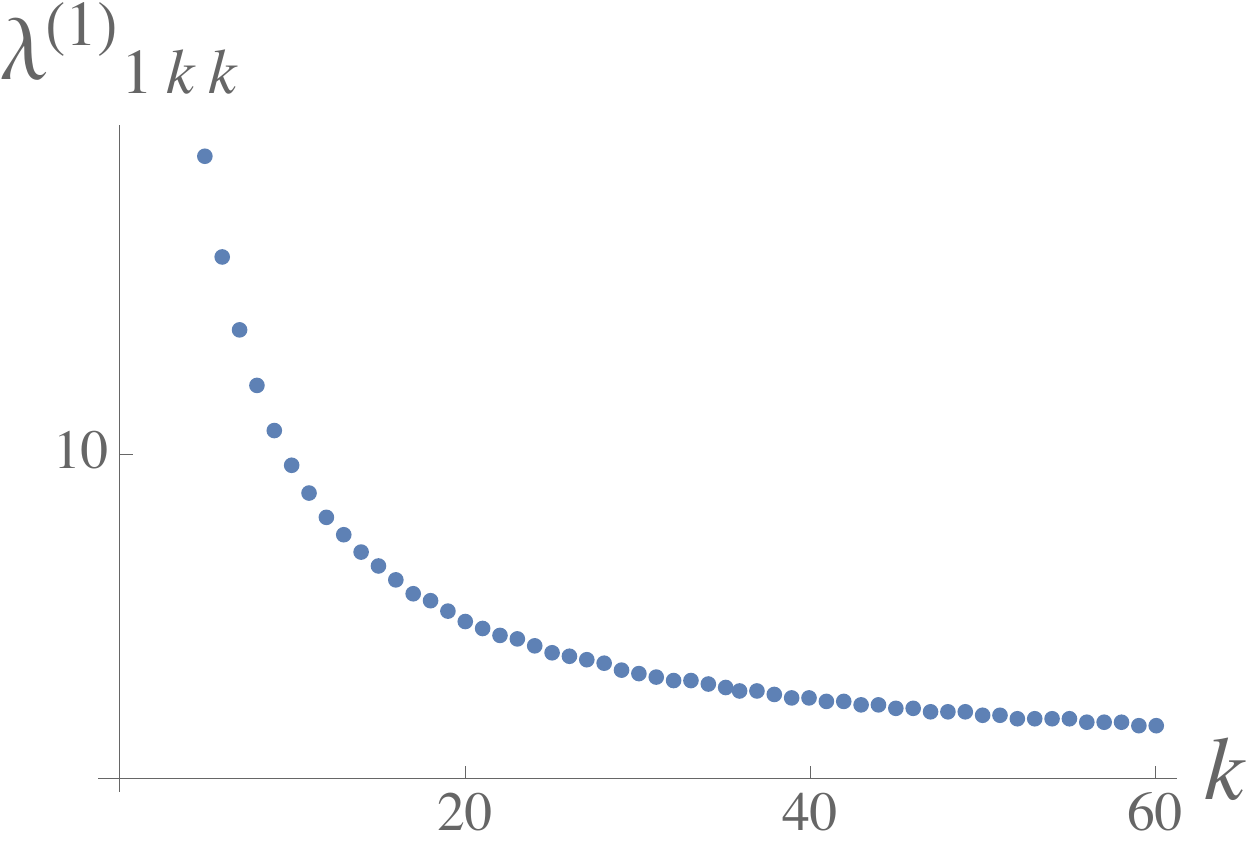}
}
\subfloat[]{
\includegraphics[width=2.1in]{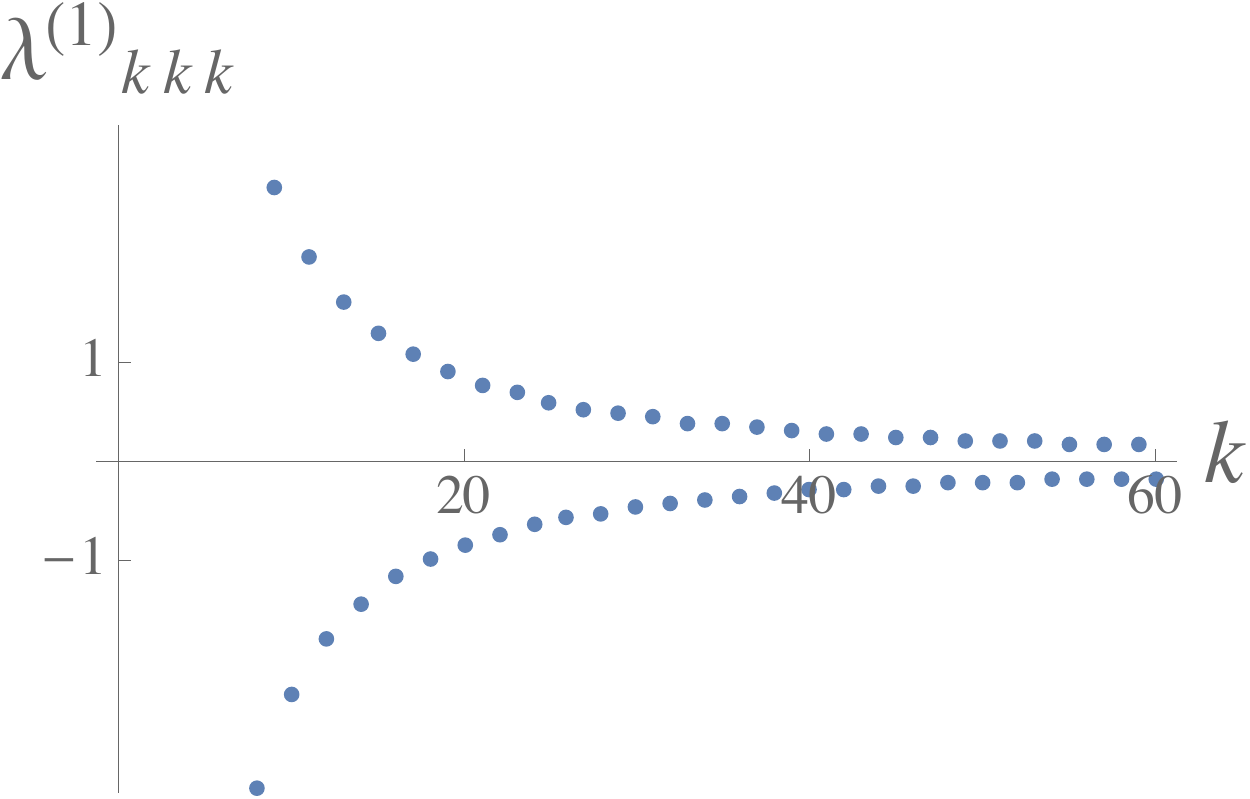}
}
\caption{The bulk cubic couplings $\lambda_{n m k }^{(1)}$ at large $q$ for some special cases.} \label{mathPlots1}
\end{figure}
In Fig.~\ref{mathPlots1} we plot $\lambda_{n m k }^{(1)}$ for a few  cases.
It is instructive to express the coupling in terms of the mass of the field. The mass of $\phi_n$ is relation to the dimension $h_n = 2n + 1 + 2\eps_n$ of $\mO_n$ through, 
\be
m_n^2 = h_n (h_n-1)= 2n(2n+1)~,
\ee
to leading order in $1/q$.
Thus, to this order, we can write, 
\be
\eps_n =  \frac{1}{q} \frac{m_n^2 +2}{m_n^2 - 2}~, \ \ \ \ \ \ \ \ \ \ \ \ \alpha_n = \sqrt{2}\, m_n  \frac{(1+4 m_n^2)^{\frac{1}{4}}}{(m_n^4 - 4)^{\frac{1}{2}}}~.
\ee
In the limit of large $n$, 
\be
\eps_n = \frac{1}{q}+ \ldots~, \  \ \ \ \ \ \ \ \ \alpha_n = \frac{2}{\sqrt{m_n}}+\ldots~, \ \ \ \ \ \ \ \ \ \ n\gg1~,
\ee
and so in the large $n$, $m$, $k$ limit the cubic coupling coming from the contact diagrams decays as, 
\be
\lambda^{(1)}_{n m k} =  - (-1)^{n + m + k}\,  \frac{384}{\sqrt{\pi}} \frac{1}{\sqrt{m_n m_m m_k}} + \ldots~, \ \ \ \ n,m, k \gg 1~.
\ee

\subsubsection{Finite $q$}
Since we were able to evaluate the contribution of the ``contact diagrams'' to the three-point function even at finite $q$ (see Eq.~\ref{I1exact}), we can write $\lambda^{(1)}_{n m k }$ at finite $q$, 
\be
\lambda_{n m k }^{(1)} = 2 \sqrt{\frac{q(q-2)}{(q-1)}\frac{\pi}{\tan \frac{\pi}{q}}}\,   A(h_n) A(h_m ) A(h_k)\,
 \Big[R(h_n, h_m, h_k) + R(h_m, h_k, h_n) + R(h_k,h_n, h_m)\Big]~,
\ee
where we defined $A(h_n)$ and $R(h_n, h_m, h_k)$ to be, 
\begin{multline}
A(h_n) = \sqrt{\frac{2 h_n - 1}{k_c'(h_n) \tan\(\pi h_n/2\)} }\frac{1}{\sin (\pi h_n)}~, \\
R(h_n, h_m, h_k) = \sin\(\frac{\pi}{2}(h_n - h_m+h_k)\) \Big(\sin(\pi (h_n + h_m)) - \sin(\pi h_m) - \sin(\pi h_k)\Big)~,
\end{multline}
where recall that $k_c'(h_n)$ is the derivative of $k_c(h)$ defined in (\ref{kch}). 

\subsection{Planar diagrams, large $q$}
\begin{figure}[t]
\centering 
\subfloat[]{\!\!\!\!\!\!\!
\includegraphics[width=2.2in]{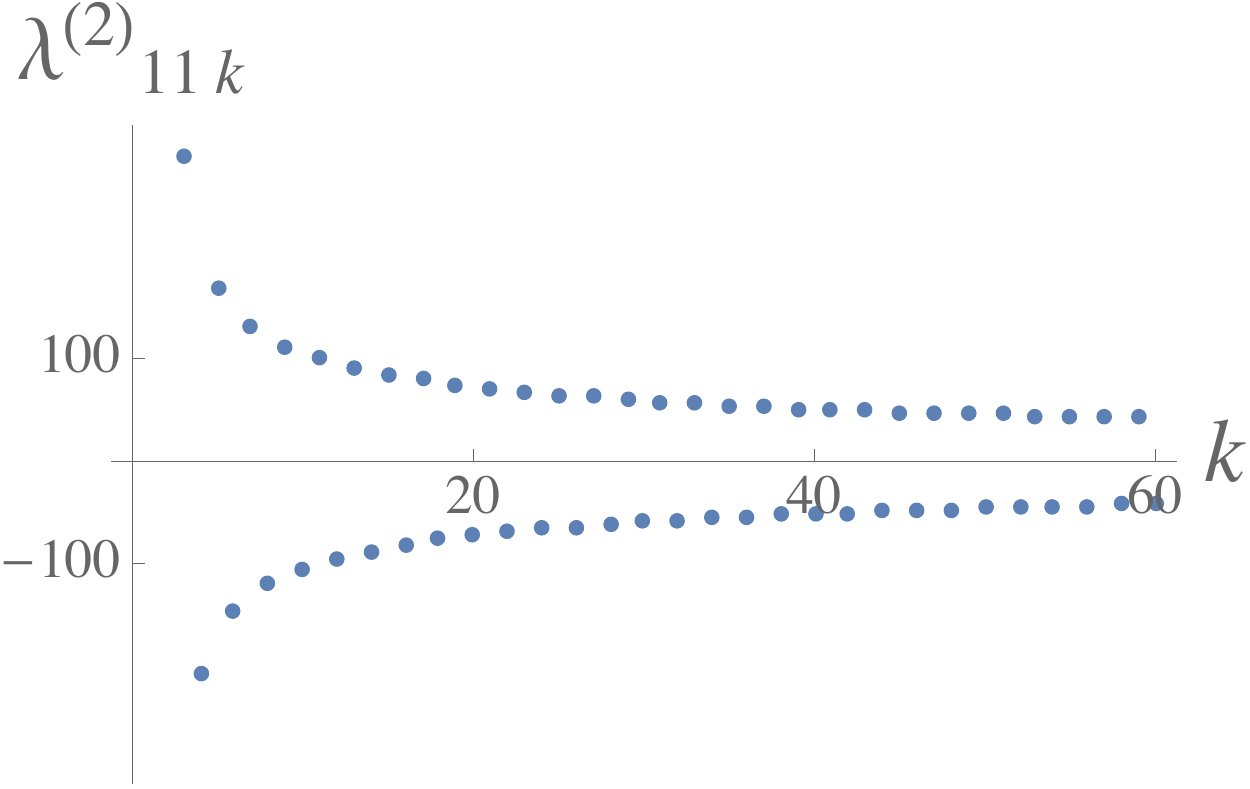}
}
\subfloat[]{
\includegraphics[width=2.1in]{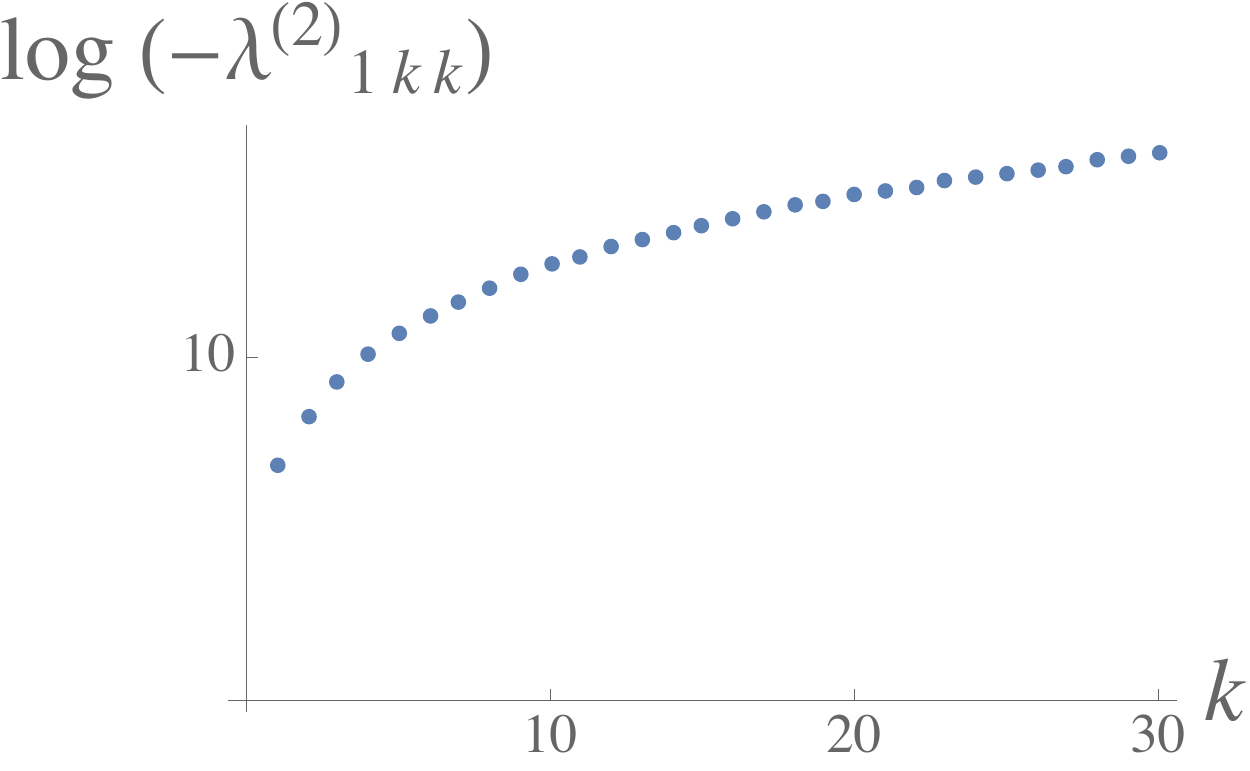}
}
\subfloat[]{
\includegraphics[width=2.1in]{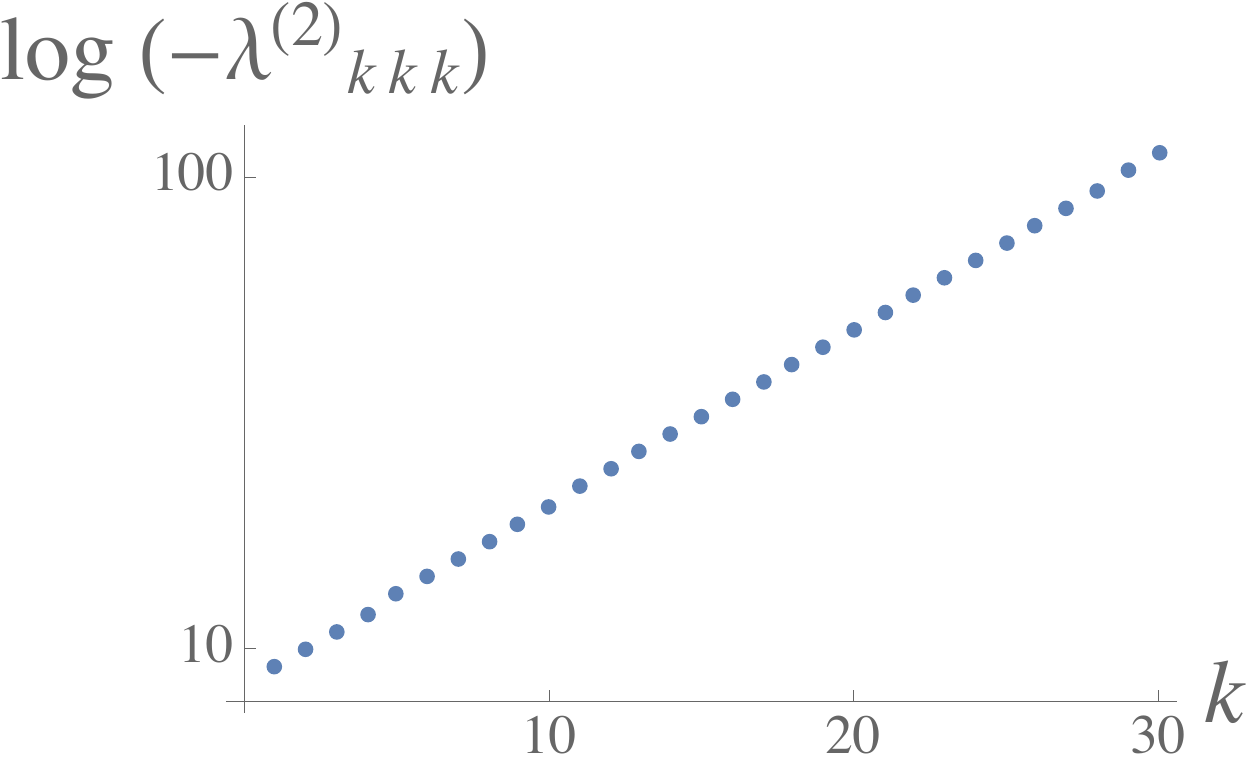}
}
\caption{The bulk cubic couplings $\lambda_{n m k }^{(2)}$ at large $q$ for some special cases. Note that (b) and (c) are log plots.} \label{mathPlots2}
\end{figure}
Now we turn to the piece of the bulk cubic coupling resulting from the planar diagrams.
Combining all the pieces from the planar diagrams, we find that it leads to the cubic coupling, 
\be \nonumber
\lambda_{n m k}^{(2)} =\! \(\frac{c_n \xi_n N_n}{\eps_n^-}\)\!\! \(\frac{c_m \xi_m N_m}{\eps_m^-}\)\!\! \(\frac{c_k \xi_k N_k}{\eps_k^-}\) \frac{s_{n m k}^{(2)}}{a_{n m k}}\[ 2\!\(\! 1 + \frac{\eps_m^-}{\eps_n^+}\)\! \! \(1 + \frac{\eps_k^-}{\eps_m^+}\)\!\! \(1 + \frac{\eps_n^-}{\eps_k^+}\)\! -\! 1 - \frac{\eps_n^- \eps_m^- \eps_k^-}{\eps_n^+ \eps_m^+\eps_k^+}\]~,
\ee
where $s_{n m k}^{(2)}$ is the triple sum discussed in Appendix~\ref{ap:sum}, and $a_{n m k}$ was given in (\ref{anmk}) and,
\be
\frac{c_n \xi_n N_n}{\eps_n^-} =\frac{\alpha_n}{2^{2n}}\frac{\Gamma(2n+2)}{\Gamma(2n+\frac{1}{2})}~,
\ee 
where $\alpha_n$ was defined in (\ref{alphan}), and recall that $\eps_n^{\pm} = \eps_n \pm \frac{1}{q}$. We have plotted  $\lambda_{n m k }^{(2)}$ for a few  cases
in Fig.~\ref{mathPlots2}.

There is a simpler and perhaps more informative way to write $\lambda_{n m k}^{(2)}$. In Appendix~\ref{ap:free} we compute the three-point function of the bilinears for a generalized free field theory of Majorana fermions in the singlet sector. We find this is related to $c_{n m k}^{(2)}$ in a simple way. Let $\lambda_{ n m k}^{free}$ denote the cubic couplings of the theory dual to this generalized free theory. Then, 
\be \nonumber
\hspace{-.5 cm} \lambda_{n m k}^{(2)} =\! \lambda_{n m k}^{free} \(\! \frac{c_n \xi_n N_n^{free}}{2  \eps_n^{-}}\! \)\!\!  \(\! \frac{c_m \xi_m N_m^{free}}{2  \eps_m^{-}}\! \)\!\! \(\! \frac{c_k \xi_k N_k^{free}}{2  \eps_k^{-}}\!\)\!\! \[ 2\!\(\! 1 + \frac{\eps_m^-}{\eps_n^+}\)\! \! \(1 + \frac{\eps_k^-}{\eps_m^+}\)\!\! \(1 + \frac{\eps_n^-}{\eps_k^+}\)\! -\! 1 - \frac{\eps_n^- \eps_m^- \eps_k^-}{\eps_n^+ \eps_m^+\eps_k^+}\]~,
\ee
where,
 \be
\frac{c_n \xi_n N_n^{free}}{2  \eps_n^{-}} = \frac{ n (2n+1)}{\sqrt{(n(1+2n)+1)(n(1+2n)-1)}}~.
\ee
In the limit of large $n, m, k$, the cubic coupling  arising the planar diagrams approaches that of the dual of the generalized free field theory, 
\be
\lambda_{n m k}^{(2)} = \lambda_{n m k}^{free} + \ldots, ~\ \ \ \ \ \ \ n, m, k \gg 1~.
\ee
For large $n, m, k$, the coupling $\lambda_{n m k }^{(2)}$ completely dominates over $\lambda_{n m k}^{(1)}$. Indeed, while $\lambda_{ k k k }^{(1)}$ has a power law decay with $k$,  $\lambda_{k k k }^{(2)}$ grows exponentially with $k$.

\section{Discussion} \label{sec:dis}

In this paper we have initiated a program of constructing the classical  (equals the large $N$ limit) bulk dual of SYK. This is a systematic procedure.  A connected $2k$-point correlation function of the fermions scales as $1/N^{k-1}$. The short time behavior of these correlation functions, in the infrared, determines the correlation functions of the ($O(N)$ singlet, large $N$ dominant) bilinear operators $\mO_n$, dual to massive scalars fields in the bulk. In this paper we studied the fermion six-point function, finding that it is given by Feynman diagrams that have three ladders glued together.  There are two  classes of such diagrams: ``contact'' diagrams in which the three ladders have a hard interaction, and planar diagrams in which the ladders connect smoothly,  pictorially reminiscent of string diagrams. From the fermion six-point function we found the three-point function of the bilinears, in terms of coefficients $c_{n m k}$.  Using the standard AdS/CFT dictionary, it was easy to write  the coefficients $\lambda_{n m k }$ of the bulk cubic interactions in terms of the $c_{n m k}$. Thus, we have determined the cubic couplings of the dual bulk theory. 

It was useful to separately analyze the two contributions to $c_{n m k}$  arising from the ``contact'' diagrams and the planar diagrams, which we denoted by $c_{n m k}^{(1)}$ and $c_{n m k}^{(2)}$, respectively, and which behave quite differently as functions of  the indices. For example, the large $n,m,k$ behavior of these coefficients is vastly different, with  $c_{n m k}^{(2)}$ much larger than  $c_{n m k}^{(1)}$  in this limit.  It would be nice to have an interpretation of each of these pieces.  

The piece, $c_{ n m k}^{(1)}$, due to the ``contact'' diagrams, is quite novel. A higher order $2 p$-point correlation function of fermions will contain an analogous contact diagram, as long as $q>p$. These diagrams would seem to give rise to an interesting kind of contribution to the bulk interaction  of the form $\lambda^{(1)}_{n_1\ldots n_p} \phi_{n_1}\cdots\phi_{n_p}$, for $p=3,4, \ldots, q-1$. If we take the limit of $q \to \infty $ (but $ q\ll N \to \infty $), then we conjecture that the bulk Lagrangian would thus, in a novel way, contain an infinite polynomial in the bulk fields, which can be calculated following the methods used in this paper.

The planar couplings, $c_{n m k}^{(2)}$, are harder to evaluate, but  reduced, in the large $q$ limit, to a finite, triple sum of products of binomial coefficients. The large $q$-limit of SYK is very interesting. In this limit,  the dimensions $h_n$ of the bilinears $\mO_n = \sum_i \chi_i \partial_{\tau}^{1+2n} \chi_i$ approach odd integers, $h_n \rightarrow 2n + 1 + O(1/q)$. In this limit the operators $O_n$ decouple from  the fermions, as the operator product coefficients of $\mO_n$ vanish. Nonetheless, the correlation functions of the $O_n$  and the bulk interactions remain finite. In this limit  we found explicit analytic expressions for $c_{n m k }$. Although we were unable, in general,  to perform the triple sums involved in evaluating  $c_{n m k}^{(2)}$, we found that these occur in a much simpler theory than SYK:   a generalized, non-local  quadratic theory of fermions in the singlet sector, with a two-point function $G_0(\tau) = \sgn(\tau)\, |\tau|^{-2\Delta}$. In Appendix~\ref{ap:free} we computed the three-point function of bilinears  in this theory, taking $\Delta$ to zero at the end, and found that its coefficient, $c_{n m k}^{free}$, is related in a simple way to $c_{ n m k}^{(2)}$. In the limit of large $n, m, k$, the two are equal. At this point, we can only speculate as to why $c_{n m k}^{free}$ and $c_{n m k}^{(2)}$ are so similar; perhaps there is a symmetry that emerges at large $q$. 

One aspect of SYK, at any $q$, is that the  bilinears $\mO_n$ do not acquire a large anomalous dimension. At large $n$ the dimensions of $\mO_n$ simply approach $2\Delta+ 2n+1$, where $\Delta$ is the IR fermion dimension, $\Delta = 1/q$. This is not surprising. Heuristically, one can think of the derivatives in $\mO_n = \sum_i \chi_i \partial_{\tau}^{1+2n} \chi_i$ as pulling the fermions apart, and so when there are many derivatives the dimension of the bilinear is just a sum of the dimensions of the constituent pieces, the two fermions and the derivatives \cite{Callan:1973pu}.~\footnote{In a gauge theory this statement is false, because of the Wilson line connecting the fermions.} Perhaps this is also the reason that we found  (at least for large $q$) that the coefficients $c_{n m k}$ of the three-point function of the bilinears approach the free field values at large $n, m, k$.

It may be interesting to study the three-point function of the bilinears in the various generalizations of SYK, of which there are many. For instance, one can add flavor to SYK \cite{GR}, leading to more refined symmetry groups and more parameters. Viewing the flavor index as a site index, one can add these flavored SYK's to form a higher-dimensional SYK \cite{Gu:2016oyy}. Or, one can consider a Lorentz invariant higher-dimensional bosonic SYK \cite{Klebanov:2016xxf}. For other generalizations, see \cite{ Fu:2016vas, Banerjee:2016ncu}. In addition, in the study of AdS$_2$/CFT$_1$ one must regulate the bulk, introducing a dilaton and turning it into ``nearly'' AdS$_2$. This is reflected in SYK by the presence of a dimension-two bilinear which breaks conformal invariance. The dilaton could in principle couple to the scalars $\phi_n$. We have focused on the purely conformal part of SYK, and so have not computed this coupling; it would be good to compute it.

\begin{figure}[t]
\centering
\subfloat[]{
\includegraphics[width=1.05in]{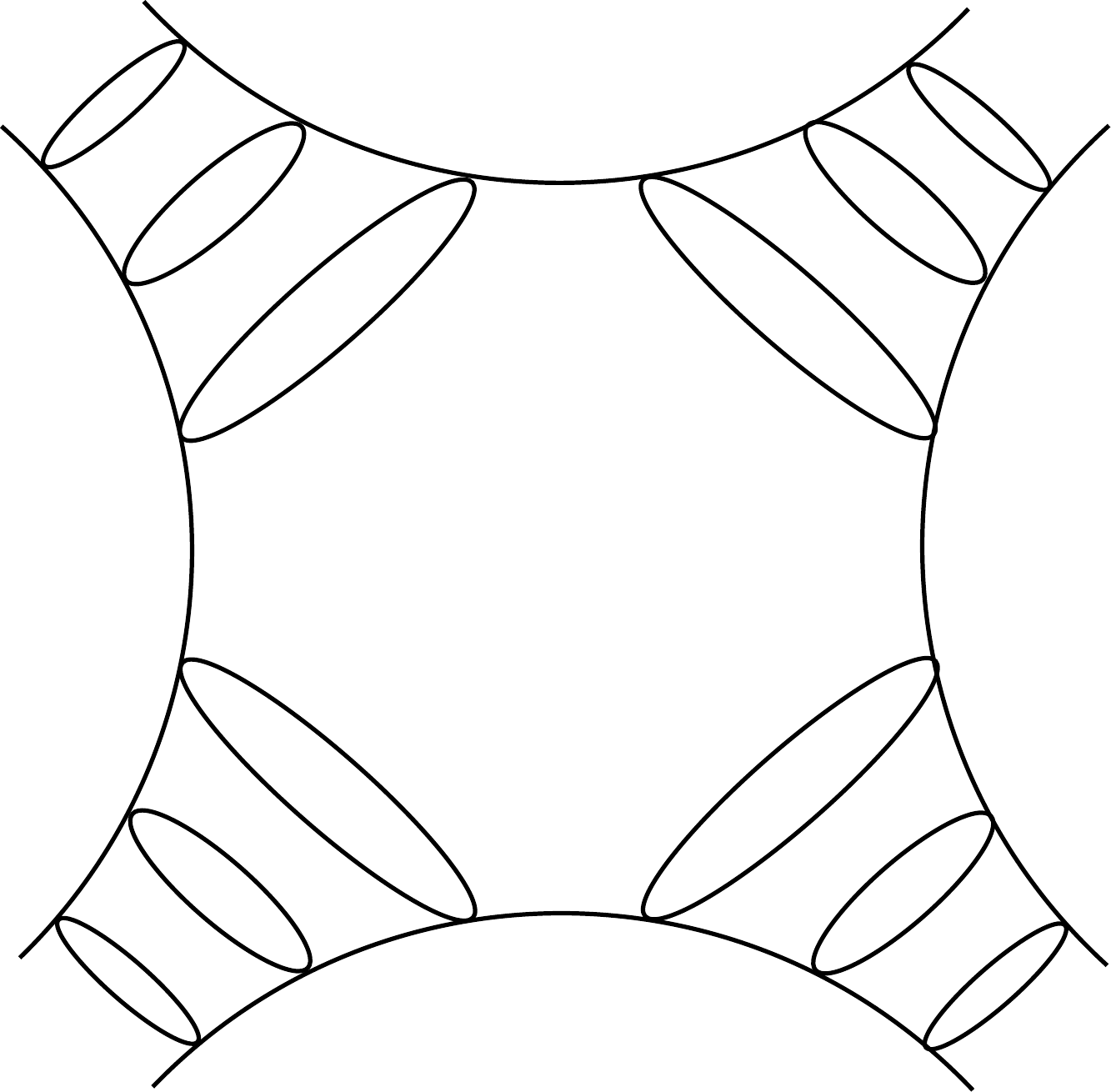}
} \ \ \ \ \ \ \ \ \ \ \ \ \ \ \ \ 
\subfloat[]{
\includegraphics[width=1.2in]{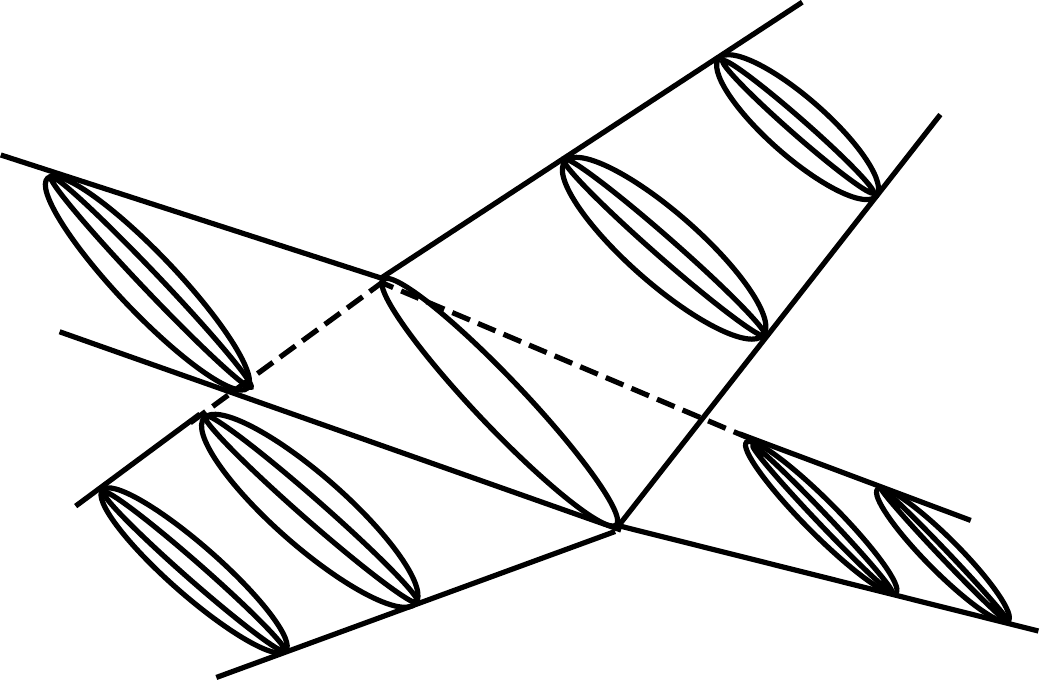}
}\\
\subfloat[]{
\includegraphics[width=1.45in]{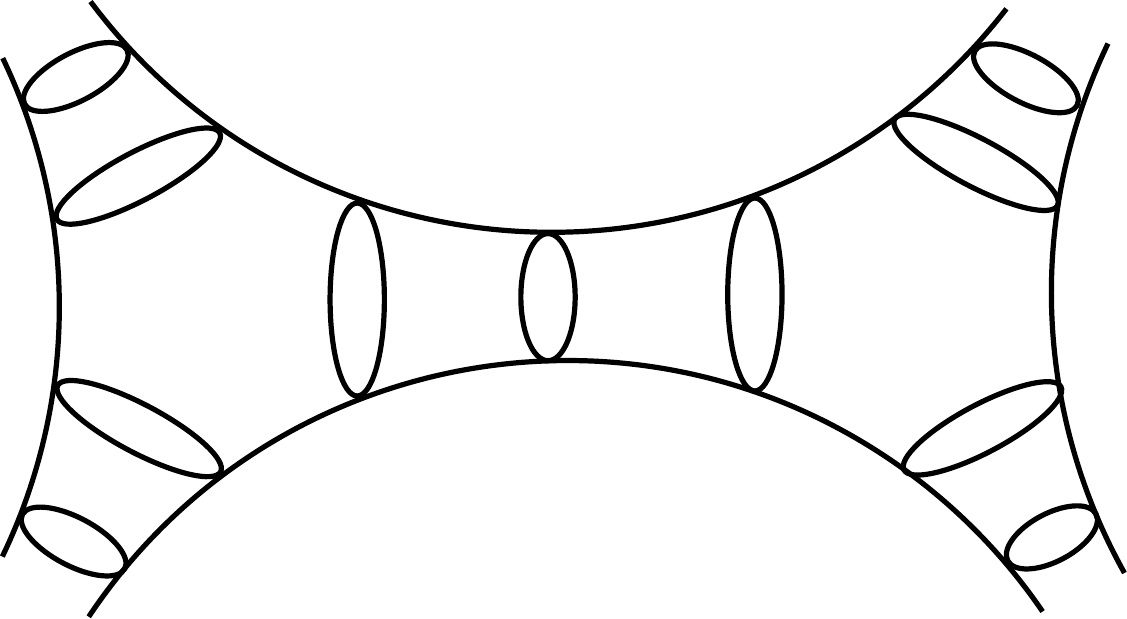}
} \ \ \ \ \ \ \ \ \ 
\subfloat[]{
\includegraphics[width=1.6in]{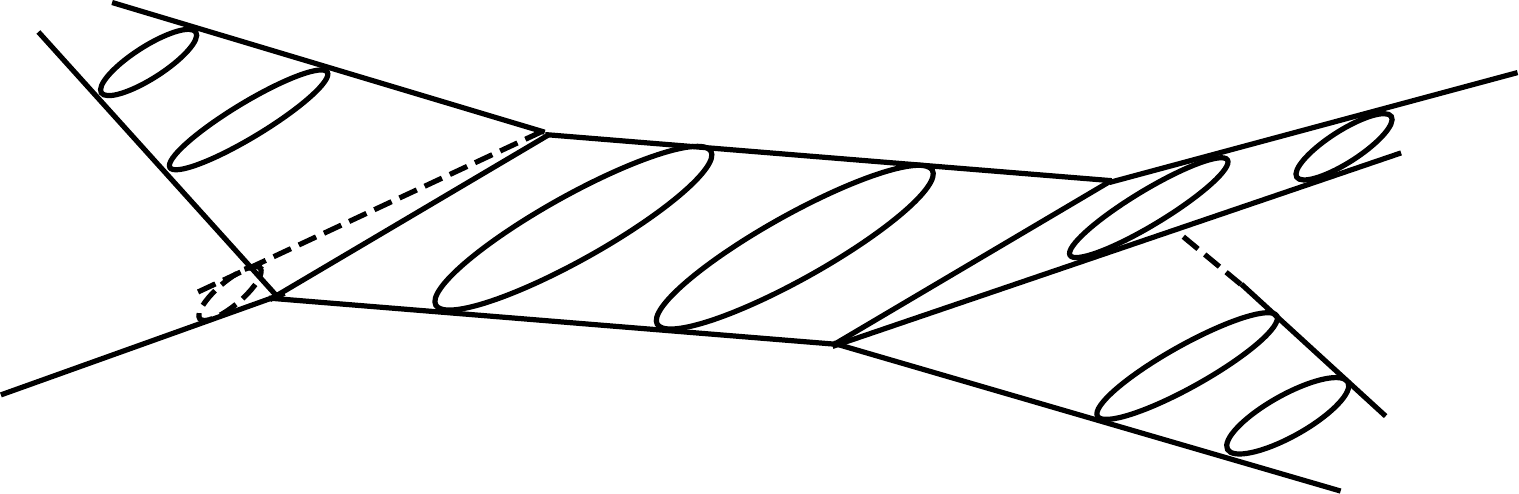}
}\ \ \ \ \ \ \ \ \ \ 
\subfloat[]{
\includegraphics[width=1.6in]{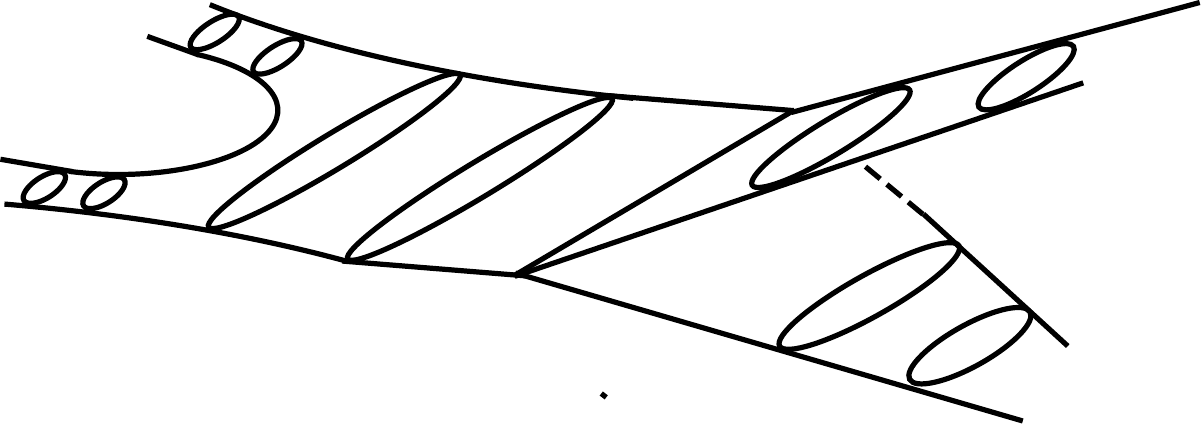}
}
\caption{Diagrams contributing to the fermion eight-point function. Note that (b) only exists for $q>4$, so we have drawn it for $q=6$. All other diagrams are for $q=4$. } \label{FigDis}
\end{figure}

In a conformal theory the operator product coefficients (together with the dimensions of  the operators) define the full content of theory, up to contact terms. We have calculated the OPE coefficients of the large $N$ dominant operators of SYK, by studying the fermion six-point function. The natural next step is to calculate the eight-point function. This will be presented in \cite{GRprog}. There are five  classes of diagrams to consider, see Fig.~\ref{FigDis}.  The first two are  simple generalizations of the diagrams relevant for the six-point function. These can be calculated using the same methods employed in this paper and  will contribute to quartic couplings in the bulk. The next three diagrams, ``exchange'' diagrams, are more interesting. In Appendix~\ref{Sec:redef} we argued that at the level of the bulk cubic interaction, as a consequence of field redefinition, one can assume that the interaction does not involve any derivatives. At the level of the quartic interaction this will no longer be the case. Matching these terms, bulk to boundary, will determine whether the quartic interactions involve derivative coupling. Indeed, an important question is whether the dual of SYK is a local quantum field theory. A Lagrangian in which every term is local, yet in which there are terms with an  arbitrarily large number of derivatives, can be nonlocal. The relevant question will be how rapidly the coefficients of the higher derivative terms decay. 

In string theory, one can also write  a Lagrangian  for the infinite number of fields, one for each mode of the string, so as to reproduce the string scattering amplitudes. Knowing that the amplitudes come from strings, that the worldsheet is the organizing principle, is far more powerful. Here too, we hope that finding the first few terms of the bulk Lagrangian will give clues towards finding the organizing principle of the bulk dual of SYK.

\bigskip

\section*{Acknowledgements} \noindent We thank J.~Bourjaily and  J.~Henn  for helpful discussions.  
This work was supported by NSF grant 1125915.

\appendix

\section{Generalized Free Fields} \label{ap:free}
In this appendix we consider the singlet sector of a generalized free field theory of Majorana fermions. We will calculate the two-point and three-point functions of the fermion bilinears.  We will find that the three-point function is the same as a factor appearing in the contribution of the planar diagrams to the large $q$ SYK three-point function of bilinears. 

We take an action,
\be
S=\frac{1}{2} \sum_{i=1}^N  \int\frac{ d\omega}{2\pi}\, \chi_i(-\omega)\, G_0(\omega)^{-1} \chi_i(\omega)~,
\ee
where we take $G_0$ to be,
\be
G_0(\omega) = i \cos(\pi \Delta) \Gamma(1-2\Delta) |\omega|^{2\Delta - 1}\, \sgn(\omega)~,\, \, \, \ \ \ \ \ \ \  \, \, G_0(\tau) = \frac{1}{2}\frac{\sgn(\tau)}{|\tau|^{2\Delta}}~.
\ee
In the limit that $\Delta\rightarrow 0$, this becomes a theory of free Majorana fermions.
It will be important for us to keep $\Delta$ finite, only taking $\Delta$ to zero at the end of the calculation.  Since the action is quadratic, it is a generalized free field theory, by which we mean that all correlation functions follow from Wick contractions. However, this theory, for finite $\Delta$, is non-local in time and it is unclear whether it has any physical meaning. This  theory is in some sense the $0+1$ dimensional analogue of the $O(N)$ model; in that case, the correlators are more involved because of spin, see for instance \cite{Sleight:2017fpc}.

The  $p$'th derivative of the two-point function is given by,
\be 
\partial_1^{p} G_0(\tau_1, \tau_2) =\frac{1}{2}\frac{\Gamma(1-2\Delta)}{\Gamma(1-2\Delta-p)} \frac{\, \, \, \sgn(\tau_{12})^{p+1}}{|\tau_{12}|^{2\Delta+ p}}~.
\ee
Taking the limit of $\Delta\rightarrow 0$, 
\be \label{Gderiv}
\partial_1^{p} G_0(\tau_1, \tau_2) =  \Delta (-1)^p (p-1)!\frac{\, \,\,  \sgn(\tau_{12})^{p+1}}{|\tau_{12}|^{p}}~.
\ee
We will be interested in the singlet-sector. 
The primaries are \cite{Craigie:1983fb}, 
\be \label{Ofree}
\mO_n =\frac{1}{\sqrt{N}} \sum_{i=1}^N \sum_{r=0}^{2n+1} d_{n r}\, \partial_{\tau}^{r}\, \chi_i\, \partial_{\tau}^{2n+1-r} \chi_i~,
\ee
where,~\footnote{The $d_{n r}$ appear to vanish for $n=0$. However, this is an artifact of normalization. If one normalizes so that the numerator is $(2n-1)!$ instead of $(2n)!$, then one will have $d_{0 0} = - d_{0 1} = 1$. }
\be \label{dnr}
d_{n r} = \frac{(-1)^r }{\Delta}\frac{(2n)!}{(2n-r)!\, (2n-r+1)!\, r!\, (r-1)!}~.
\ee
Proceeding to the two-point function of the primaries, by Wick contraction we have, 
\be \nonumber
\!\!\langle \mO_n(\tau_2) \mO_m(\tau_4)\rangle\! =\!\underset{\tau_3 \rightarrow \tau_4}{\underset{\tau_1 \rightarrow \tau_2}{\text{lim}}}\sum_{r, s}\! d_{n r} d_{m s}  \partial_1^{r}\, \partial_{2}^{2n+1\!-r}\,  \partial_3^{s}\, \partial_4^{2m+1\!-s} \Big[G_0(\tau_1, \tau_4) G_0(\tau_2, \tau_3)-G_0(\tau_1, \tau_3)  G_0(\tau_2, \tau_4) \Big].
\ee
Evaluating the derivatives using (\ref{Gderiv}) gives,
\begin{multline}
\langle \mO_n(\tau_2) \mO_m(\tau_4)\rangle = \frac{1}{|\tau_{24}|^{2(n+m+1)}}
 \sum_{r=0}^{2n+1} \sum_{s=0}^{2m+1} d_{n r} d_{m s} \Big[ (r+s-1)! (2n+2m+1-r-s)! \Big.\\ \Big.  - (2n-r+s)!(2m+r-s)!\Big]~.
\end{multline}
Explicitly performing the sums, we get for the bilinear two-point function, 
\be
\langle \mO_n(\tau_2) \mO_m(\tau_4)\rangle =(N_n^{free})^2 \frac{  \delta_{n m}}{|\tau_{24}|^{4n+2}}~, \ \ \ \, \, \, \, \, \,(N_n^{free})^2 =  \frac{2^{4n+1}}{(2n+1)  }\frac{\Gamma(2n+\frac{1}{2})}{\sqrt{\pi}\,  \Gamma(2n)}~.
\ee
For the three-point function, again employing Wick contractions, we get, 
\begin{multline} \nonumber
\langle \mO_n(\tau_2) \mO_m(\tau_4)\mO_k (\tau_6) \rangle =\underset{\tau_5\rightarrow \tau_6}{\text{lim}}\underset{\tau_3 \rightarrow \tau_4}{\text{lim}}\underset{\tau_1 \rightarrow \tau_2}{\text{lim}}\, \frac{1}{\sqrt{N}}\sum_{r=0}^{2n+1}\sum_{s=0}^{2m+1} \sum_{t=0}^{2k+1} d_{n r} d_{m s} d_{k t}\, \partial_1^r\, \partial_2^{2n+1-r}\\ \partial_3^s\, \partial_4^{2m+1-s}\, \partial_5^{t}\, \partial_6^{2k+1-t}
\Big[ G(\tau_1, \tau_6) G(\tau_2, \tau_3) G(\tau_4, \tau_5) + \text{perm}\Big]~,
\end{multline}
where the permutations involve interchanging $\tau_1$ with $\tau_2$, or interchanging $\tau_3$ with $\tau_4$, or interchanging $\tau_5$ with $\tau_6$, for a total of eight terms. All eight terms will give the same contribution, so we have,
\begin{multline} \nonumber
\langle \mO_n(\tau_2) \mO_m(\tau_4)\mO_k (\tau_6) \rangle 
=N^{-\frac{1}{2}}\, 8 \Delta^3 (2n)! (2m)! (2k)!\, \sgn(\tau_{24} \tau_{26}\tau_{46})\\
\sum_{r=0}^{2n+1} \sum_{s=0}^{2m+1} \sum_{t=0}^{2k+1} d_{n r} d_{m s} d_{k t} (2n-r+s)! (2k+r-t)! (2m-s+t)! \frac{(-1)^s }{\tau_{24}^{2n+1-r+s} \tau_{26}^{2k+1+r-t}\tau_{46}^{2m+1+t-s}}~.
\end{multline}
In fact, this sum is familiar. It is essentially $s_{n m k}^{(2)}$ (\ref{mCSum}) encountered in computing the sum of the planar diagrams contributing to the three-point function of the primaries in SYK for large $q$. Some properties of this sum are discussed in Appendix~\ref{ap:sum}. In any case, we have for the three-point function in this generalized free theory, 
\be \label{Free3pt}
\langle \mO_n(\tau_2) \mO_m(\tau_4) \mO_k(\tau_6) \rangle =\frac{1}{\sqrt{N}}\frac{ c_{n m k}^{free}}{|\tau_{24}|^{h_{n} +h_{m}  - h_{k}}| \tau_{46}|^{h_{m} + h_{k} - h_{n}} |\tau_{62}|^{h_{k} + h_{n} - h_{m}}}
\ee
where $h_n = 2n +1$ and,
\be
 c_{n m k}^{free} = 8\, s_{n m k}^{(2)}~.
\ee
\subsection*{Summary}
It is convenient to renormalize the operators so that the two-point function is order one, 
\be
\langle \mO_n(\tau_2) \mO_m(\tau_4)\rangle =\frac{  \delta_{n m}}{|\tau_{24}|^{2 h_n}}~,
\ee
and then the conformal three-point function (\ref{Free3pt}) has the coefficient,  
\be
 c_{n m k}^{free} =  \frac{8\, s_{n m k}^{(2)}}{N_n^{free} N_m^{free} N_k^{free} }~.
\ee
This is the result for three-point function of bilinears in this particular generalized free field theory. It is similar to the piece of the three-point function of  SYK that comes from summing planar diagrams. The coefficient of the three-point function that we found there was, 
 \be
 c_{n m k }^{(2)} = c_n \xi_n\, c_m \xi_m\, c_k\xi_k\, s_{n m k}^{(2)} 
 %\frac{\eps_n^+ + \eps_m^-}{\eps_n^+ \eps_m^-}\frac{\eps_m^+ + \eps_k^-}{\eps_m^+ \eps_k^-}\frac{\eps_k^+ + \eps_n^-}{\eps_k^+ \eps_n^-}~.
 \( 2 \frac{\eps_n^+ + \eps_m^-}{\eps_n^+ \eps_m^-}\frac{\eps_m^+ + \eps_k^-}{\eps_m^+ \eps_k^-}\frac{\eps_k^+ + \eps_n^-}{\eps_k^+ \eps_n^-} - \frac{1}{\eps_n^+\eps_m^+\eps_k^+} - \frac{1}{\eps_n^- \eps_m^-\eps_k^-}\)~.
 \ee
The ratio of the two is,
\be \label{ratio} \nonumber
 \frac{c_{n m k }^{(2)}}{ c_{n m k}^{free} } =\!\! \(\frac{c_n \xi_n N_n^{free}}{2  \eps_n^{-}}\) \!\! \(\frac{c_m \xi_m N_m^{free}}{2 \eps_m^{-}}\)\!\! \(\frac{c_k \xi_k N_k^{free}}{2  \eps_k^{-}}\) \!
 %\(1 + \frac{\eps_m^-}{\eps_n^+}\)\(1 + \frac{\eps_k^-}{\eps_m^+}\)\(1 + \frac{\eps_n^-}{\eps_k^+}\)~,
 \[ 2\!\(\! 1 + \frac{\eps_m^-}{\eps_n^+}\)\! \! \(1 + \frac{\eps_k^-}{\eps_m^+}\)\!\! \(1 + \frac{\eps_n^-}{\eps_k^+}\)\! -\! 1 - \frac{\eps_n^- \eps_m^- \eps_k^-}{\eps_n^+ \eps_m^+\eps_k^+}\]~,
 \ee
 where
 \be
\frac{c_n \xi_n N_n^{free}}{2  \eps_n^{-}} = \frac{ n (2n+1)}{\sqrt{(n(1+2n)+1)(n(1+2n)-1)}}~.
\ee

\section{Integrals} \label{ap:integrals}
In this appendix we collect a few useful integrals.
We start with two-point integrals. Defining $ a= a_1 + a_2$, 
\bea \label{2pt}
&&\int d\tau_0 \frac{1}{|\tau_{01}|^{ 2 a_1}} \frac{1}{|\tau_{02}|^{2 a_2}} 
= \pi^{\frac{1}{2}} \frac{\Gamma\(\frac{1}{2}-a_1\)}{\Gamma(a_1)}\frac{ \Gamma\(\frac{1}{2}-a_2\) }{ \Gamma(a_2)}\frac{\Gamma\(a-\frac{1}{2}\)}{ \Gamma\(1 - a\)}    \frac{1}{|\tau_{12}|^ {2a-1}}~, \\\label{2pt1}
&&\int d\tau_0 \frac{\sgn(\tau_{01})}{|\tau_{01}|^{2 a_1}} \frac{1}{|\tau_{02}|^{2 a_2}}  = -\pi^{\frac{1}{2}}\frac{\Gamma\(1-a_1\)}{\Gamma(\frac{1}{2}+a_1)}\frac{ \Gamma\(\frac{1}{2}-a_2\) }{ \Gamma(a_2)}\frac{\Gamma\(a\)}{ \Gamma\(\frac{3}{2} - a\)}    \frac{\sgn(\tau_{12})}{|\tau_{1 2}|^{2a-1}}~,  \\ \label{2Int}
&&\int d\tau_0 \frac{\sgn(\tau_{01})}{|\tau_{01}|^{2 a_1}} \frac{\sgn(\tau_{02})}{|\tau_{02}|^{2a_2}} =\pi^{\frac{1}{2}} \frac{\Gamma\(1-a_1\)}{\Gamma(\frac{1}{2}+ a_1)}\frac{ \Gamma\(1-a_2\) }{ \Gamma(\frac{1}{2}+a_2)}\frac{\Gamma\(a-\frac{1}{2}\)}{ \Gamma\(1 - a\)}    \frac{1}{|\tau_{1 2}|^{2 a-1}}~,
\eea
These can be obtained, for instance, by Fourier transforming both sides. 

Now consider a three-point integral.  If $a_1+a_2 + a_3 = 1$ then,
\be \label{A4}
\int d\tau_0 \frac{1}{|\tau_{01}|^{2 a_1}}\frac{1}{|\tau_{02}|^{2 a_2}} \frac{1}{|\tau_{03}|^{2 a_3}} = \pi^{\frac{1}{2}} \frac{\Gamma\(\frac{1}{2}\!-a_1\)}{\Gamma(a_1)}  \frac{\Gamma\(\frac{1}{2}\!-a_2\)}{\Gamma(a_2)} \frac{\Gamma\(\frac{1}{2}\!-a_3\)}{\Gamma(a_3)} \frac{1}{|\tau_{12}|^{1 - 2 a_3}|\tau_{13}|^{1-2 a_2} |\tau_{23}|^{1 - 2 a_1}}~.
\ee
A standard way to evaluate such an integral is by introduction of Schwinger parameters. A faster way is by noticing that the integral transforms as a conformal three-point function, thereby fixing the functional form on the right-hand side of (\ref{A4}). The constant is then fixed by taking $\tau_3$ to infinity and using (\ref{2pt}). This same method allows us to find, for $a_1 + a_2 + a_3=1$,
\be \label{cf3ptint}
\!\!\! \int\! d\tau_0 \frac{\sgn(\tau_{01})}{|\tau_{01}|^{2 a_1}}\frac{\sgn(\tau_{02})}{|\tau_{02}|^{2 a_2}} \frac{1}{|\tau_{03}|^{2 a_3}} = \pi^{\frac{1}{2}} \frac{\Gamma\(1\!-a_1\)}{\Gamma\(\frac{1}{2}+ a_1\)}  \frac{\Gamma\(1\!-a_2\)}{\Gamma\(\frac{1}{2}+a_2\)} \frac{\Gamma\(\frac{1}{2}\!-a_3\)}{\Gamma\(a_3\)} \frac{\sgn(\tau_{13} \tau_{23})}{|\tau_{12}|^{1 - 2 a_3}|\tau_{13}|^{1-2 a_2} |\tau_{23}|^{1 - 2 a_1}}~.
\ee

\subsection{Selberg Integral}
The Selberg integral is an $n$ dimensional integral defined as, 
\be
S_{n, n}(\alpha, \beta, \gamma)=\int_{[0,1]^n}d \tau_1\cdots d \tau_n \prod_{i=1}^{n} |\tau_i|^{\alpha-1} |1-\tau_i|^{\beta-1} \prod_{1\leq i<j\leq n} |\tau_{i j}|^{2\gamma}~.
\ee
The integral is given by, 
\be
S_{n,n}(\alpha, \beta, \gamma) = \prod_{j=0}^{n-1} \frac{\Gamma(\alpha + j \gamma) \Gamma(\beta+ j \gamma) \Gamma(1+(j+1)\gamma)}{\Gamma(\alpha+\beta + (n+j-1)\gamma) \Gamma(1+\gamma)}~.
\ee
We will need a generalization of the Selberg integral, denoted by $S_{n, p}$~,
\be
S_{n, p}(\alpha, \beta, \gamma) = \int_{[0,1]^p} \int_{[1,\infty)^{n-p}} d \tau_1 \cdots d \tau_n\, \prod_{i=1}^{n} t_i^{\alpha-1} |1-\tau_i|^{\beta-1} \prod_{1\leq i<j\leq n} |\tau_{i j}|^{2\gamma}~,
\ee
for $0\leq p \leq n$. 
This has the same integrand as the Selberg integral, but the integration region is now over $1<\tau_i<\infty$ for $n-p$ of the times $\tau_i$. The Selberg integral is a special case, $S_{n, n}(\alpha, \beta, \gamma)$. Through an inversion of the integration variables $\tau_i \rightarrow 1/\tau_i$, one trivially gets the relation, 
\be
S_{n, p } (\alpha, \beta, \gamma) = S_{n, n-p}\(1-\alpha-\beta -2(n-1) \gamma, \beta, \gamma\)~.
\ee
A more involved relation between the different $S_{n, p}$ is \cite{Forrester, Dotsenko:1984ad}, 
\be
S_{n, p} (\alpha, \beta, \gamma) = S_{n, p-1}(\alpha, \beta, \gamma) \frac{ p}{n-p+1} \frac{\sin\( \pi (n-p+1)\gamma\) \sin\(\pi(\alpha + \beta + (n+p-2) \gamma\)}{\sin(\pi p \gamma) \sin\(\pi (\alpha + (p-1) \gamma)\)}~.
\ee
Successively applying this relation allows one to express all $S_{n, p}$ in terms of the Selberg integral $S_{n, n}(\alpha, \beta, \gamma)$. 

The integral we will be interested in has two integration variables, with the integration domain being the entire plane, 
\be
S_{2, 2}^{full}(\alpha, \beta, \gamma) = \int_{-\infty}^{\infty} d\tau_1 \int_{-\infty}^{\infty} d\tau_2\,\, |\tau_1|^{\alpha-1} |\tau_2|^{\alpha-1} |1-\tau_1|^{\beta-1} |1-\tau_2|^{\beta-1}|\tau_{12}|^{2\gamma} ~.
\ee
Breaking up the integration into different regions and, through an appropriate change of variables, relating the integral in each to $S_{2,p}$, we find,
\begin{multline} \label{SelFull}
S_{2, 2}^{full}(\alpha, \beta, \gamma) = S_{2,2}(\alpha, \beta, \gamma)+ S_{2, 2}(1-\alpha- \beta - 2 \gamma, \beta, \gamma) + S_{2,2}(1-\alpha -\beta- 2\gamma,\alpha, \gamma) \\+ 2\, S_{2,1}(\alpha, \beta, \gamma) + 2\, S_{2,1} (1-\alpha - \beta - 2\gamma, \alpha, \gamma) +2\, S_{2,1}(\alpha, 1 - \alpha-\beta - 2 \gamma, \gamma)~.
\end{multline}

\section{Triple Sum} \label{ap:sum}
In this appendix we discuss some properties of the triple sum encountered in the computation of the three-point function of bilinears. This sum was encountered both in SYK, arising from the planar diagrams at large $q$ as discussed in Sec.~\ref{sec:I2largeq}, as well as in free field theory discussed in Appendix~\ref{ap:free}. The sum (\ref{mCSum}) was given by, 
\be \label{mCSuma}
\!\!\!\!\!s_{n m k}^{(2)} = -\!\! \sum_{p_1,p_2, p_3} \!\binom{2n}{p_1}\!\binom{2m}{p_2}\!\binom{2 k}{p_3}\!
 \binom{2n\!+\!p_2\!-\!p_1}{p_2+1}\!\binom{2m\!+\!p_3\!-\!p_2}{p_3+1}\!\binom{2 k \!+\!p_1\!-\!p_3}{p_1+1}\!
 \frac{z^{p_1-p_2 + 2m - 2 k}}{ (\!-1\!-\!z)^{p_3 - p_2+2n - 2 k}} ~.
 \ee
 As noted before, it is independent of $z$. 
For arbitrary $n,m$ and small fixed $k$, $s_{n m k}^{(2)}$ takes a simple form, 
\bea
s_{n m 1}^{(2)} &=& 2 \frac{\Gamma(2n+2m-1)}{\Gamma(2n) \Gamma(2m)} \Big( n (2n+1) + m(2m+1) -1\Big)~,\\ \nonumber
s_{n m 2}^{(2)} &=& \frac{4}{3} \frac{ \Gamma( 2n + 2 m - 3)}{\Gamma(2n) \Gamma(2m)}\Big( \[n(1+2n)\(27 - 10 n(1+2n) +n^2 (1+2n)^2\)\] + \[n\rightarrow m\] \Big. \\ \nonumber
&\,&\, \, \, \, \, \,\, \, \, \,   \qquad \qquad\qquad - 18 + 6nm(1+2n)(1+2m) ( n(1+2n) + m(1+2m) - 5) \Big.\Big)~.
\eea
As $k$ is increased, $s_{n m k}^{(2)}$ looks progressively more complicated. Let us now look at $s_{n m k}^{(2)}$ for $n=m=k$. For low values of $n$ one has, 
\be
\begin{tabular}{c | c c c c c  }
$n$ & $1$ & $2$ & $3$ & $4$ & $5$ \\
\hline
$s_{n n n}^{(2)}$ & $20$ & $8464$ & $4050864$ & $2116980800$ & $1173644492800$
\end{tabular}
\ee
Clearly, $s_{n n n}^{(2)}$ does not take a simple form. In fact, series A181544 in the encyclopedia of integers sequences \cite{oesc} is defined as, 
\be
t(p, q) = \[x^{q}\] (1-x)^{3p+1} \sum_{r=0}^{p+q} \binom{p+r}{r}^3 x^r~,
\ee
where $\[x^{q}\]$ denotes the coefficient of $x^q$ of what follows after it. One can see that, 
\be
s_{n n n }^{(2)} = t(2n, 2n-1)~.
\ee
Looking at general $s_{n m k }^{(2)}$, one can analyze the sum (\ref{mCSuma}) by finding its recursion relations. It is more natural to view (\ref{mCSuma}) with $n, m, k$ allowed to take half-integer values. Defining $F(n, m, k) = s_{\frac{n}{2} \frac{m}{2} \frac{k}{2}}^{(2)}$, we find two recursion relations,
\bea \nonumber
(n-k)F(n, m, k) + (n-k)F(n,m-1,k) + n F(n-1,m,k) - k F(n, m,k-1) &= &0~, \\ \nonumber
(m-k)F(n,m,k) +(m+n-k)F(n,m-1,k) -(n+k-m)F(n,m,k-1)& =&0~.
\eea
These recursion relations by themselves are not sufficient to fix $s_{n m k}^{(2)}$, however they do allow us to express $s_{n m k}^{(2)}$ in terms of $s_{p q 1}^{(2)}$ and $s_{r r r }^{(2)}$. This is unlikely to take a form significantly simpler than (\ref{mCSuma}), so we will not pursue this further. 
A final comment is that we know the generating function for this sum $s_{n m k}^{(2)}$. It is $f^{2}_{n m k}$ given in (\ref{f20}). In other words, $s_{n m k}^{(2)}$ is equal to (\ref{smethod1}).

\section{Field Redefinition} \label{Sec:redef}
The goal of this paper has been to derive the cubic couplings of the bulk dual of SYK by computing the three-point function of the bilinears in SYK, and then applying the AdS/CFT dictionary. In translating the CFT result into a statement about the bulk cubic couplings, in Sec.~\ref{sec:bulk} we assumed that the action takes the form,
\be 
S_{bulk} = \int d^2 x\sqrt{g}\[ \frac{1}{2} (\nabla\phi_n)^2 + \frac{1}{2} m_n^2 \phi_n^2 +\frac{1}{\sqrt{N}} \, \lambda_{n m k}\,  \phi_n \phi_m \phi_k\]~.
\ee
 However,  one could have also considered an action in which the cubic terms contain derivatives. From the computation of the CFT three-point function, we would be unable to tell if the bulk coupling does or does not have derivatives.~\footnote{The distinction would show up at the level of the four-point function.}   However, as we show in this appendix,  at this order in $1/N$ a cubic coupling with derivatives is equivalent to one without derivatives. Going between the two is simply a matter of field redefinition. 

In particular, consider a possible term in the bulk Lagrangian of the form, 
\be \label{gen3}
\mathcal{L}_{deriv} = f(\nabla) \phi_1\,  g(\nabla)  \nabla_{\mu} \phi_2\,  h(\nabla) \nabla^{\mu} \phi_3~,
\ee
where $f,g,h$ are some string of derivatives, acting on $\phi_1, \phi_2, \phi_3$, respectively. The overall expression is of course a scalar. We can integrate by parts to rewrite this as, 
\be \label{redef}
\!\mathcal{L}_{deriv}\! =\! \frac{1}{2}\Big(\! f(\nabla) \nabla^2\phi_1\,\,   g(\nabla) \phi_2\,\, h(\nabla) \phi_3 - f(\nabla) \phi_1\,\, g(\nabla) \nabla^2  \phi_2\,\, h(\nabla) \phi_3 
-f(\nabla) \phi_1\,\, g(\nabla) \phi_2\,\, h(\nabla)\nabla^2 \phi_3\!\Big)~.
\ee
If we make the field redefinition, 
\be \label{phi1re}
\phi_1 \rightarrow \phi_1 + \frac{(-1)^l }{2} f(\nabla)\( g(\nabla) \phi_2 h(\nabla) \phi_3\)~,
\ee
where $l$ is the number of derivatives in $f(\nabla)$, then  the kinetic term becomes, 
\be
(\nabla \phi_1)^2 \rightarrow (\nabla \phi_1)^2  - f(\nabla) \nabla^2 \phi_1\, g(\nabla) \phi_2\, h(\nabla) \phi_3+ \ldots~,
\ee
where we have integrated by parts. We thus precisely cancel off the first term in (\ref{redef}). An analogous redefinition of $\phi_2$ and $\phi_3$ takes care of the second and third terms in (\ref{redef}). Under the mapping (\ref{phi1re}) the mass term becomes, 
\be
m_1^2 \phi_1^2 \rightarrow m_1^2 \phi_1^2 + \frac{1}{2} m_1^2 f(\nabla) \phi_1\,   g(\nabla) \phi_2\, h(\nabla) \phi_3~.
\ee
This is a piece of our new cubic interaction.  Compared to (\ref{gen3}), our redefinition has removed two of the derivatives. One can continue this procedure until all derivatives are removed.

%\bibliographystyle{utphys}
%\bibliography{6ptbib}

\bibliographystyle{utphys}
%\bibliography{GRbib}

\end{document}